\documentclass[12pt,a4paper]{article}

\usepackage{amsmath}
\usepackage{amssymb}
\usepackage{bbm}
\usepackage{dirtytalk}
\usepackage{braket}
\usepackage[style=numeric-comp, sorting=none, bibstyle=ieee, url=false]{biblatex}
\bibliography{biblio}
\numberwithin{equation}{section}

\usepackage{subcaption}
\usepackage{adjustbox}

\usepackage{tikz}
\usepackage[colorlinks, allcolors=blue]{hyperref}
\usepackage{booktabs}
\usepackage{multirow}

\usetikzlibrary{decorations.markings}
\tikzset{->-/.style={decoration={markings,
  mark=at position #1 with {\arrow{>}}},postaction={decorate}}}
\tikzset{-<-/.style={decoration={markings,
  mark=at position #1 with {\arrow{<}}},postaction={decorate}}}

\newcommand{\Z}{{\mathbb{Z}}}

\newcommand{\C}{{\mathbb{C}}}

\newcommand{\pqty}[1]{\left( #1 \right)}
\newcommand{\bqty}[1]{\left[ #1 \right]}
\newcommand{\abs}[1]{\left\lvert #1\right\rvert}

\newcommand{\expval}[1]{\langle #1 \rangle}

\newcommand {\dv}[3][ ]{
  \ifx #1 { }
    \frac{d #2}{d #3}
  \else
    \frac{d^{#1} #2}{d #3^{#1}}
  \fi
}

\newcommand{\tr}{\operatorname{tr}}

\renewcommand{\Re}{\operatorname{Re}}

\newcommand{\Or}{\mathrm{O}}

\newcommand{\SU}{\mathrm{SU}}
\newcommand{\U}{\mathrm{U}}

\title{Hamiltonians and gauge-invariant Hilbert\\ space for lattice Yang-Mills-like theories\\ with finite gauge group}

\author{A.\ Mariani$^1$, S.\ Pradhan$^2$, E.\ Ercolessi$^2$}

\date{%
    $^1$ \small{Albert Einstein Center for Fundamental Physics, Institute for Theoretical Physics,
University of Bern, Sidlerstrasse 5, CH-3012 Bern, Switzerland}\\[2ex]%
    $^2$ \small{Dipartimento di Fisica, Universit\`a di Bologna and INFN, Via Irnerio 46, 40126 Bologna, Italy}\\%
}

\begin{document} 

\maketitle

\begin{abstract} \normalsize

Motivated by quantum simulation, we consider lattice Hamiltonians for Yang-Mills gauge theories with finite gauge group, for example a finite subgroup of a compact Lie group.
We show that the electric Hamiltonian admits an interpretation as a certain natural, non-unique Laplacian operator on the finite Abelian or non-Abelian group, and derive some consequences from this fact.
Independently of the chosen Hamiltonian, we provide a full explicit description of the physical, gauge-invariant Hilbert space for pure gauge theories and derive a simple formula to compute its dimension.
We illustrate the use of the gauge-invariant basis to diagonalize a dihedral gauge theory on a small periodic lattice.

\end{abstract}

\newpage
 
\section{Introduction}

Quantum simulation is a field of growing interest both experimentally and theoretically \cite{Feynman, buluta2009simulators, cirac2012goals, georgescu2014simulation}.
It holds the promise to overcome technical difficulties, such as the \textit{sign problem} \cite{Troyer_Wiese, condmattsignproblem, banuls2020lgtreview}, which affect numerical Monte Carlo simulations in several interesting regimes, with applications to particle physics and condensed matter systems \cite{AartsReview, condmattsignproblem}.
In recent years, advances in experimental techniques have allowed the proposal and realisation of different setups to simulate physical theories via quantum mechanical systems \cite{quantumsimulation,actual_simulation, blatt2012trappedions, bloch2008ultracold, bloch2012ultracold, aspuruguzik2012photonic, houck2012superconducting, angelakis2017quantum, ultracold, singha2011lattice, DigitalSim1, DigitalSim2}.
The long-term hope is that experimental and theoretical advances may one day allow the quantum simulation of some currently inaccessible non-perturbative aspects of strongly coupled theories.

A particularly interesting application is the quantum simulation of gauge theories \cite{MegaReview, banuls2020lgtreview, dalmonte2016lgtreview, wiese2013ultracold, ZoharReview, hauke2013trappedions, marcos2013sclgt}, which are ubiquitous in both particle and condensed matter physics.
From a theoretical standpoint, quantum simulation is most natural in a Hamiltonian formulation with a finite-dimensional Hilbert space.
One possibility to achieve this is by replacing the gauge group, a Lie group, with a finite group, for example one of its finite subgroups \cite{Ercoetal1, Ercoetal2, ZoharBurrello}.
Finite subgroups of Lie groups were already considered in the early days of lattice field theory \cite{Rebbi,BhanotCreutz,SU3Subgroups,SU2Subgroups,ZNgeneralizedAction,SU2Subgroups2}.
It was found that, unlike Lie groups, finite groups undergo a \say{freezing} transition, beyond which they cease to approximate the physics of their parent Lie group.
It was later pointed out, however, that as long as an appropriate comparison is made in the vicinity of the phase transition, finite group gauge theory based on sufficiently large, but fixed, subgroups of the Lie group provides an effective description of the Lie group gauge theory \cite{Hasenfratz1}.
Gauge theories with finite gauge group have also found direct applications in condensed matter physics \cite{DualGaugeTheory2D,DualGaugeTheory3D,CondMatDiscreteGaugeTheory2,Manjunath_2021} and quantum gravity \cite{TransferMatrixFiniteGroup}.
In the Hamiltonian formulation, the Abelian case was considered in \cite{Ercoetal1, Ercoetal2, SchwingerCiracBanuls} while the general formulation in the non-Abelian case was given in \cite{ZoharBurrello}.
This formulation, however, lacked an explicit form of the electric term of the Hamiltonian.
In \cite{TransferMatrixFiniteGroup} the full Hamiltonian for a finite group gauge theory was derived via the transfer-matrix formulation from the Wilson action. 

In the present work we consider a class of finite-group gauge theories in the Hamiltonian formulation, which mimic some aspects of Lie group Yang-Mills theories on the lattice.
The Hamiltonian is based on the construction of a natural Laplacian operator on the finite group, and is valid for any Abelian or non-Abelian finite group.
As a special case, this includes both the finite-group Hamiltonian obtained via the transfer-matrix formulation from the finite-group Wilson action, but also a wider class of non-Lorentz invariant theories.
The characterization of the Hamiltonian using the finite-group Laplacian may be used to obtain non-trivial physical information about the theory.
Irrespective of the choice of Hamiltonian, we show that spin-network states are particularly suitable to give a description of the physical, gauge-invariant Hilbert space for pure gauge theories, and, based on this, we derive a simple formula to compute the dimension of the physical Hilbert space.
Finally, we illustrate the use of the gauge-invariant basis by constructing the Hamiltonian for a gauge theory based on the dihedral group and compute some quantities of interest via exact diagonalization.

\section{Finite group gauge theory}\label{sec:hamiltonian formulation}

\subsection{The Hilbert space}

\subsubsection{Basic construction}\label{sec:basic hilbert space}

In the Hamiltonian formulation of lattice gauge theories \cite{KogSuss, Osborne, ZoharBurrello}, time is continuous while the $d$ spatial dimensions are discretized into a hypercubic lattice.
Classically, we assign a group element $g \in G$ to each spatial lattice link, where $G$ is the gauge group.
In the Lie group case, one would typically write $U_\mu(x) = \exp{(iA_\mu(x))} \in G$ for the gauge field variable assigned to the lattice link between points $x$ and $x + \hat{\mu}$, where $A_\mu(x)$ is the vector potential.
Links are oriented, and if a link is traversed in the opposite orientation, then $g$ is replaced with $g^{-1}$.
Note that finite groups have no Lie algebras, so we work with group-valued quantities as far as possible.
In what follows, we write $g \in G$ for a group element indifferently for both finite and Lie groups $G$.

\medskip

Since a classical configuration is given by a choice of group element $g$ on each lattice link, in the quantum theory the states in the Hilbert space of each link are generally given by a superposition \cite{Osborne}
\begin{equation}
    \ket{\psi}=\int dg \, \psi(g)\ket{g} \ ,
\end{equation}
where $\{\ket{g}\}$ is the group element orthonormal basis, consisting of one state $\ket{g}$ per group element $g$; it can be thought of as a \say{position basis} on the group.
In the case of a Lie group, the wavefunction $\psi(g)$ is square-integrable with respect to the Haar measure.
Hence, the Hilbert space on each link can then be identified with $L^2(G)$, i.e. the space of square-integrable functions on $G$ \cite{Osborne}.
For a finite group, the Haar measure is replaced with a sum over the group elements, $\int dg = \sum_g$, and the Hilbert space is simply the \textit{group algebra} $\C[G]$, which is the complex vector space spanned by the group element basis.

The overall Hilbert space is then given by the tensor product $\mathcal{H}_{\text{tot}} = \bigotimes_{\mathrm{links}} L^2(G)$, or $\mathcal{H}_{\text{tot}} = \bigotimes_{\mathrm{links}} \C[G]$.
Note that for a finite group, $\C[G]$ has finite dimension, because it is spanned by the finitely-many group element states $\{\ket{g}\}$.
Therefore the Hilbert space on each link is finite-dimensional and $\mathcal{H}$ is finite-dimensional on a finite lattice; the dimension is given by $\abs{G}^L$ where $L$ is the number of lattice links.
For a Lie group, on the other hand, we have infinitely many basis states $\{\ket{g}\}$ and therefore the Hilbert space is infinite-dimensional \textit{on each link}.

\medskip

In the Hamiltonian formulation of gauge theories, the statement that the theory is invariant under gauge transformations translates at the level of the Hilbert space by restricting the allowed states only to those which are gauge-invariant.
In particular, on the single-link Hilbert space one can define left and right \say{translation} operators, in the analogy where $\{\ket{g}\}$ is a position basis in group space \cite{ZoharBurrello},
\begin{equation}
    \label{eq:regular representations}
    L_g \ket{h} = \ket{gh} \ , \quad \quad R_g \ket{h} = \ket{hg^{-1}} \ .
\end{equation}
A local gauge transformation is given by a choice of group element $g_x \in G$ at every site $x$ of the lattice \cite{Osborne}.
This acts on the overall Hilbert space $\mathcal{H}_{\text{tot}}$ via the operator
\begin{equation}
    \label{eq:gauss law operator}
    \mathcal{G}(\{g_x\}) = \bigotimes_{l=\expval{xy} \in \mathrm{links}} L_{g_{x}} R_{g_y} \ ,
\end{equation}
where $\{g_x\}$ is an arbitrary choice of group elements $g_x$ at each lattice site $x$, and the link $l$ connects the points $x$ and $y$.
In other words, each link state $\ket{g_l}$ transforms as $\ket{g_l} \mapsto \ket{g_x g_l g_y^{-1}}$.

The only physical states are those which satisfy the so-called \say{Gauss' law} constraint \cite{KogSuss, Osborne, Tong}
\begin{equation}
    \label{eq:lattice gauss law}
    \mathcal{G}(\{g_x\}) \ket{\psi}=\ket{\psi} \ ,
\end{equation}
for any possible choice of local assignments $\{g_x\}$ of group variables to lattice sites.
The Gauss law \eqref{eq:lattice gauss law} is an exponentiated version of the usual Gauss law formulated in terms of Lie algebra generators. 
The states which satisfy \eqref{eq:lattice gauss law} form the physical, gauge-invariant Hilbert space $\mathcal{H}_\mathrm{phys}$.
Note that the condition \eqref{eq:lattice gauss law} only involves group-valued quantities and is thus valid for both Lie groups and finite groups.
In the case of finite groups, the condition simplifies because it is sufficient to impose invariance against a set of generators of the finite group.

One can also straightforwardly include matter fields such as fermion fields which live on each lattice site.
Under a gauge transformation, they transform as $\Psi(x) \to R(g_x) \Psi(x)$, where $R$ is some representation of the gauge group.

\subsubsection{The representation basis and the Peter-Weyl theorem}\label{sec:representation basis}

It turns out to be fruitful to introduce a different basis of the overall Hilbert space $\mathcal{H}$, \say{dual} to the group element basis.
The operators $L_g$ and $R_g$ introduced in \eqref{eq:regular representations} are unitary representations of $G$, known as the \textit{left} and \textit{right regular} representations \cite{Serre, KnappLieGroups}.
This is because $L_g L_h = L_{gh}$ and $(L_g)^{-1}=L_{g^{-1}}=(L_g)^\dagger$, as can be explicitly checked by acting on the group element basis, and the same holds for $R$.
Their representation theory leads to the Peter-Weyl theorem \cite{KnappLieGroups, Osborne, marianithesis}, which states that for a finite or compact Lie group $G$,
\begin{equation}
    \label{eq:peterweyl}
    L^2(G) = \bigoplus_{j \in \Sigma} V_j^* \otimes V_j \ ,
\end{equation}
where $j$ is a label for the irreducible representations (irreps) of $G$, and $\Sigma$ is the set of all irreps of $G$.
For finite groups $L^2(G)$ is replaced, as usual, with $\C[G]$.
Here $V_j$ is the representation vector space corresponding to representation $j$ and $V_j^*$ its dual vector space.
For both compact Lie groups and finite groups the irreps are finite-dimensional and can be chosen to be unitary.
For a finite group, $\Sigma$ is a finite set, while it is countably infinite for a compact Lie group \cite{KnappLieGroups, Serre}.
In terms of the Peter-Weyl decomposition, the left and right regular representations take a particularly simple form \cite{marianithesis},
\begin{equation}
    \label{eq:translations peter weyl}
    L_g R_h = \bigoplus_j \rho_j(g)^* \otimes \rho_j(h) \ ,
\end{equation}
where $\rho_j$ is the matrix of the $j$th irrep of $G$.
The individual action of either $L_g$ or $R_h$ may be obtained by setting either $g$ or $h$ to the identity.
Eq.~\eqref{eq:translations peter weyl} is especially useful because, as we will see in Section \ref{sec:physical Hilbert space}, it simplifies the action of the Gauss' law constraint \eqref{eq:lattice gauss law}.

\medskip

The Peter-Weyl theorem provides an alternative basis for the single-link Hilbert space.
For each irrep $j$ one chooses appropriate bases for $V_j^*$ and $V_j$, which we denote $\{\ket{jm}\}$ and $\{\ket{jn}\}$ respectively, where $1 \leq m,n \leq \dim{j}$.
Here $\dim{j} \equiv \dim{V_j}$ is the dimension of the representation.
On each representation subspace, we use the shorthand notation $\ket{jmn} \equiv \ket{jm} \otimes \ket{jn}$.
Then the \say{representation basis} for $L^2(G)$ or $\C[G]$ is given by the set $\{\ket{jmn}\}$ for all $j \in \Sigma$ and $1 \leq m,n \leq \dim{j}$.
In terms of the group element basis, one has \cite{ZoharBurrello}
\begin{equation}
    \label{eq:change of basis}
    \expval{g|jmn}= \sqrt{\frac{\mathrm{dim}(j)}{\abs{G}}} \left[\rho_j(g)\right]_{mn} \ ,
\end{equation}
where the bases $\{\ket{jm}\}$, $\{\ket{jn}\}$ are chosen so that $\rho_j$ is unitary.
It should be emphasized that \eqref{eq:change of basis} is valid for both finite and compact Lie groups; $\abs{G}$ is either the order of the finite group or the volume $\abs{G}\equiv\int dU\, 1$ given by the possibly unnormalized Haar measure \cite{Osborne, marianithesis}.
It is a basic result of the representation theory of finite groups that $\sum_j \pqty{\dim{j}}^2 = \abs{G}$, which ensures that the group element basis and the representation basis have the same number of states \cite{Serre}.

Since every group admits a trivial, one-dimensional irrep with $\rho(g)\equiv 1$, we always have a singlet representation state $\ket{0}$, which may be extended to the whole lattice to form the \say{electric ground state} $\ket{0_E}$,
\begin{equation}
    \label{eq:electric ground state}
    \ket{0_E} = \bigotimes_{l \in \mathrm{links}} \ket{0}_l \ , \quad \quad \ket{0}=\frac{1}{\sqrt{\abs{G}}} \sum_g \ket{g} \ ,
\end{equation}
where we used\eqref{eq:change of basis} to express $\ket{0}$ in the group element basis.
We have summarized the representation theory of some groups of interest in Appendix \ref{sec:some groups of interest}.

\medskip

In the specific case of the group $\Z_N$, the representations are all one-dimensional because $\Z_N$ is Abelian and therefore $m=n=1$ and can be omitted.
If $\xi$ is a generator of $\Z_N$, then the group elements are $\Z_N = \{1, \xi, \xi^2, \ldots, \xi^{N-1}\}$ and the irreps are simply $\rho_j (\xi^k) = \omega_N^{kj}$ for $j = 0,1,\ldots, N-1$, with $\omega_N= e^{2\pi i /N}$.
The bases $\{\ket{\xi^k}\}$ and $\{\ket{j}\}$ are related by
\begin{equation}
    \ket{j} = \sum_{k=0}^{N-1} \braket{\xi^k|j} \ket{\xi^k} = \frac{1}{\sqrt{N}}\sum_{k=0}^{N-1} \omega_N^{kj} \ket{\xi^k} \ ,
\end{equation}
which is just the discrete Fourier transform.
In the case of the dihedral group $D_4$ with $8$ elements, we have four one-dimensional representations, each of which spans a one-dimensional subspace of $\C[G]$.
We then have a two-dimensional representation which spans a $2^2=4$ dimensional subspace of $\C[G]$ through the four basis elements $\ket{jmn}$ for $1 \leq m, n \leq 2$.

\subsection{The Hamiltonian}

\subsubsection{Basic construction}\label{sec:hamiltonian basic construction}

The Hamiltonian for a Yang-Mills gauge theory on the lattice takes the form \cite{KogSuss, ZoharBurrello,TransferMatrixFiniteGroup, Osborne, Caspar_Wiese}
\begin{equation}
    \label{eq:generic hamiltonian}
    H = \lambda_E \sum_{l \in \mathrm{links}} h_E + \lambda_B \sum_{\square} h_B(g_\square) \ ,
\end{equation}
where $h_E$ depends only on each lattice link, while $h_B$ depends on the lattice plaquettes $\square$ and $g_\square = g_1 g_2 g_3^{-1} g_4^{-1}$ is the product of the four link variables in a lattice plaquette with the appropriate orientation.
It is also possible to add matter fields, but we focus here on the pure gauge theory.

If the gauge group is a Lie group, each group element $g = e^{iX}$ can be written as the exponential of a Lie algebra element $X$.
Then one also has infinitesimal generators of left-translations $\hat{\ell}^a_L$ such that $L_{e^{iT^a}} = \exp{\pqty{i \hat{\ell}_L^a}}$, where $T^a$ is a Lie algebra basis and $a$ a color index \cite{Osborne}.
In other words, $\hat{\ell}_L$ is the Lie algebra representation corresponding to the group representation $L$, and plays the role of the chromoelectric field.

The Lie group Hamiltonian, also known as the Kogut-Susskind Hamiltonian, is then given by \cite{KogSuss, Osborne}
\begin{equation}
    \label{eq:kog suss hE hB}
    h_E = \sum_{a} \pqty{\hat{\ell}^a_L}^2 \ , \quad \quad \quad \quad h_B = 2(\dim{\rho} - \Re \tr{\rho(g_\square)}) \ ,
\end{equation}
where $\rho$ is the fundamental representation of $\SU(N)$, with couplings $\lambda_E = g^2/2$ and $\lambda_B = 1/g^2$ in terms of a coupling constant $g$ (the lattice spacing is set to $1$). As the group element basis may be thought of as a \say{position basis} in group space, the infinitesimal generators of translations $\hat{\ell}_L^a$ may be thought of as \say{momentum} operators in group space.
Then the electric Hamiltonian $h_E$, which is the sum of the squares of the \say{momenta} in all directions, is a Laplacian in group space.
Applying the Peter-Weyl decomposition \eqref{eq:translations peter weyl} to $\hat{\ell}_L^a$, one finds that \cite{Osborne,marianithesis}
\begin{equation}
    \label{eq:laplacian decomposition}
    h_E = \sum_{a} \pqty{\hat{\ell}^a_L}^2 = \sum_{jmn} C(j) \ket{jmn}\bra{jmn} \ ,
\end{equation}
where $C(j)$ is the quadratic Casimir operator, which depends only on the representation $C(j)$.
For $\U(1)$, for example $C(j)=j^2$, while for $\SU(2)$ one finds $C(j) = j(j+1)$.

\medskip

We note that the magnetic Hamiltonian depends only on group-valued quantities and is therefore well-defined for both Lie groups and finite groups.
On the other hand, the electric Hamiltonian depends on the infinitesimal Lie algebra through $\hat{\ell}_L^a$ and therefore the definition does not extend to finite groups.
The decomposition \eqref{eq:laplacian decomposition} is well-defined also for finite groups, but one must leave the coefficients $C(j)$ unsatisfactorily unspecified because finite groups do not have a Casimir operator \cite{ZoharBurrello}.

If one thinks of a finite group as a natural discretization of some parent Lie group, the natural choice of electric Hamiltonian is a discrete Laplacian on the finite group.
The geometric structure of a finite group is that of a graph, with group elements as vertices and the group operation defining the edges.
This is called a \textit{Cayley graph}.
The discrete Laplacian on the finite group is then naturally given by the graph Laplacian of the Cayley graph.
This choice also preserves the interpretation of the electric Hamiltonian as a quantum-mechanical rotor in group space \cite{KogSuss}.

We explain the construction of the finite group Laplacian in detail in Section \ref{sec:finitegrouplaplacian}, and the resulting Hamiltonian takes the form of \eqref{eq:generic hamiltonian} with 
\begin{equation}
    \label{eq:generalized ym hamiltonian}
    \begin{gathered}
        H = \lambda_E \sum_{l \in \mathrm{links}} h_E + \lambda_B \sum_{\square} h_B(g_\square) \ , \\
        h_E = \sum_{g \in \Gamma} (1-L_g)  \ , \quad \quad \quad \quad h_B= h_B (g_\square) \ ,
    \end{gathered}
\end{equation}
where $\Gamma \subset G$ is a subset of the group (\textit{not} a subgroup) such that
\begin{enumerate}
    \item $1 \not \in \Gamma$, i.e.
$\Gamma$ doesn't contain the identity element.
    \item $\Gamma^{-1}=\Gamma$, i.e.
it is invariant under inversion of group elements.
In other words, if $g \in \Gamma$, then $g^{-1} \in \Gamma$ also.
    \item $g \Gamma g^{-1} = \Gamma$, i.e.
it is invariant under conjugation.
In other words, $\Gamma$ is a union of conjugacy classes of $G$.
\end{enumerate}
These conditions on $\Gamma$ ensure that the electric Hamiltonian is gauge-invariant.
On the other hand, as usual, the magnetic term is gauge-invariant as long as $h_B$ is any real function such that $h_B(g_1 g_\square g_1^{-1})=h_B(g_\square)$ for any $g_1 \in G$.
As explained in Section \ref{sec:action formulation}, the Hamiltonian \eqref{eq:generalized ym hamiltonian} includes as a special case the transfer-matrix Hamiltonian obtained in \cite{TransferMatrixFiniteGroup} which consists in a certain specific choice of subset $\Gamma$.
The choice of $\Gamma$ is in fact not unique, a fact which we will also discuss in later sections.

\medskip

While the magnetic Hamiltonian $h_B$ is diagonal in the group element basis, the electric Hamiltonian $h_E$ is diagonal in the representation basis, and in fact \cite{ZoharBurrello}
\begin{equation}
    \label{eq:electric hamiltonian rep basis}
    h_E = \sum_{jmn} h_E(j) \ket{jmn}\bra{jmn} \ , \quad \quad \quad h_E(j)=\abs{\Gamma} - \frac{1}{\dim{j}} \sum_{g \in \Gamma} \chi_j(g)\ ,
\end{equation}
where $\abs{\Gamma}$ is the number of elements in $\Gamma$ and $\chi_j$ is the character of the irrep labelled $j$.
The electric Hamiltonian may be interpreted as an \say{on-link} hopping term within group space; in fact, up to a constant, it may be written as 
\begin{equation}
    h_E = -\sum_{g \in \Gamma} \sum_{h \in G} \ket{gh}\bra{h} \ ,
\end{equation}
 and it favours each link to sit in the electric ground state \eqref{eq:electric ground state}, which is fully delocalized in group space.
On the other hand, the magnetic term is a plaquette-based potential which pushes plaquettes close to the identity.
The competition between the two non-commuting terms gives rise to non-trivial dynamics.

\medskip

We would like to emphasize that the description of the electric Hamiltonian $h_E$ in \eqref{eq:generalized ym hamiltonian} as the graph Laplacian of the Cayley graph associated with the group is not simply an interesting analogy, but also a tool which may be used to extract information on the Hamiltonian itself.
As an example, we note the well-known fact that the smallest eigenvalue of a graph Laplacian is always zero (given by the trivial representation state \eqref{eq:electric ground state}) and its degeneracy equals the number of connected components of the graph \cite{spectralgraphtheory}.
Moreover, it is not hard to show that if $\Gamma$ does not generate the group $G$, but rather only a subgroup $\expval{\Gamma} < G$, then the Cayley graph splits into connected components which are identified with the cosets of $\expval{\Gamma}$ in $G$.
The number of such components, and therefore the degeneracy of the ground state of $h_E$ on each link, is given by
\begin{equation}
    \mathrm{electric\,ground\,state\,degeneracy}=\frac{\abs{G}}{\abs{\expval{\Gamma}}} \ .
\end{equation}
This is the degeneracy of $h_E$ on each link; the degeneracy of the electric Hamiltonian $H_E = \sum_{\mathrm{links}} h_E$ on the Hilbert space of the whole lattice is larger.
If, instead, $\Gamma$ generates the whole group, then the electric Hamiltonian is not degenerate.
A detailed proof can be found in Appendix \ref{sec:laplacian degeneracy}.
The degeneracy of the electric ground state is not only an important feature of the theory, but also technically important for methods such as adiabatic quantum simulation.

As we will see at the end of Section \ref{sec:finitegrouplaplacian}, an electric Hamiltonian with degenerate ground state can be constructed in the simple case of the dihedral group $D_4$.
In general, the electric ground state degeneracy can also occur with the choice of $\Gamma$ arising from the transfer-matrix formulation of the Wilson action, as described in Section \ref{sec:action formulation}.
For example, consider the permutation group $G=S_5$. Starting from the Wilson action in the six-dimensional faithful irrep of $S_5$, one finds $h_E$ with $\Gamma$ equal to the conjugacy class of the $5$-cycles; then $\Gamma$ generates the subgroup $\expval{\Gamma} = A_5$ and since $\abs{S_5}/\abs{A_5}=2$, the electric Hamiltonian ground state is two-fold degenerate, with the ground states spanned by the two representation states corresponding to the one-dimensional irreps.

\subsubsection{The finite group Laplacian}\label{sec:finitegrouplaplacian}

\begin{figure}
    \centering
    \subfloat[]{
        \centering
        \begin{tikzpicture}
    \node[shape=circle,draw=black, minimum width=1cm] (g0) at (0,0) {$e$};
    \node[shape=circle,draw=black, minimum width=1cm] (g1) at (2,0) {$g$};
    \node[shape=circle,draw=black, minimum width=1cm] (g2) at (2.5,1.65)     {$g^2$};
    \node[shape=circle,draw=black, minimum width=1cm] (g3) at (1,2.65)     {$g^3$};
    \node[shape=circle,draw=black, minimum width=1cm] (g4) at (-0.5,1.65)     {$g^4$};

    \path [-] (g0) edge (g1);
    \path [-] (g1) edge (g2);
    \path [-] (g2) edge (g3);
    \path [-] (g3) edge (g4);
    \path [-] (g4) edge (g0);
\end{tikzpicture}
    }
    \hfill
    \subfloat[]{
        \centering
        \begin{tikzpicture}
    \node[shape=circle,draw=black, minimum width=1cm] (g0) at (0,0) {$e$};
    \node[shape=circle,draw=black, minimum width=1cm] (g1) at (2,0) {$g$};
    \node[shape=circle,draw=black, minimum width=1cm] (g2) at (2.5,1.65)     {$g^2$};
    \node[shape=circle,draw=black, minimum width=1cm] (g3) at (1,2.65)     {$g^3$};
    \node[shape=circle,draw=black, minimum width=1cm] (g4) at (-0.5,1.65)     {$g^4$};
    
    \path [-] (g0) edge (g1);
    \path [-] (g0) edge (g2);
    \path [-] (g1) edge (g2);
    \path [-] (g1) edge (g3);
    \path [-] (g2) edge (g3);
    \path [-] (g2) edge (g4);
    \path [-] (g3) edge (g4);
    \path [-] (g3) edge (g0);
    \path [-] (g4) edge (g0);
    \path [-] (g4) edge (g1);
\end{tikzpicture}    
    }
    \hfill
    \subfloat[]{
        \centering
        \begin{tikzpicture}
    \node[shape=circle,draw=black, minimum width=1cm] (e) at (0,0) {$e$};
    \node[shape=circle,draw=black, minimum width=1cm] (r) at (1.5,0)     {$r$};
    \node[shape=circle,draw=black, minimum width=1cm] (r2) at (3,0) {$r^2$};
    \node[shape=circle,draw=black, minimum width=1cm] (r3) at (4.5,0)     {$r^3$};
    \node[shape=circle,draw=black, minimum width=1cm] (s) at (0,1.5) {$s$};
    \node[shape=circle,draw=black, minimum width=1cm] (rs) at (1.5,1.5)     {$rs$};
    \node[shape=circle,draw=black, minimum width=1cm] (r2s) at (3,1.5) {$r^2s$};
    \node[shape=circle,draw=black, minimum width=1cm] (r3s) at (4.5,1.5)     {$r^3s$};
    
    \path [-] (e) edge (r);
    \path [-] (e) edge (s);
    
    \path [-] (e) edge[bend right=45] (r3);
    \path [-] (s) edge[bend left=45] (r3s);
    
    \path [-] (r) edge (r2);
    \path [-] (r) edge (rs);
    
    \path [-] (r2) edge (r3);
    \path [-] (r2) edge (r2s);
    
    \path [-] (r3) edge (r3s);
    \path [-] (s) edge (rs);
    
    \path [-] (rs) edge (r2s);
    \path [-] (r2s) edge (r3s);
\end{tikzpicture}
    }
    \hfill
    \caption{Examples of Cayley graphs. (a) and (b) show $\Z_5$ with $\Gamma = \{g,g^{-1}\}$ and $\Gamma = \{g,g^2, g^{-1}, g^{-2}\}$ respectively. (c) shows $D_4$ with $\Gamma = \{r,r^{-1}, s\}$ }
    \label{fig:examples of Cayley graphs}
\end{figure}
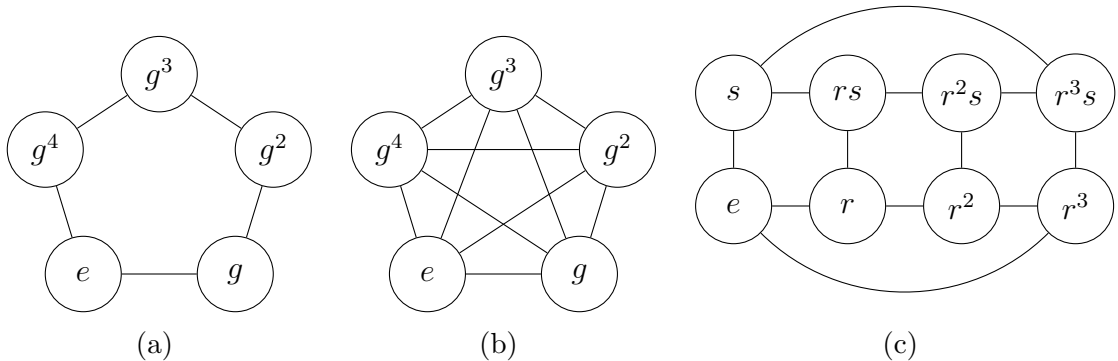

In this section we explain in detail the construction of the electric Hamiltonian on each link as the finite group Laplacian, i.e. the graph Laplacian of the Cayley graph of the finite group.
Given a finite group $G$, we choose a set of generators $\Gamma \subset G$, which we require to be invariant under inversion, that is $\Gamma^{-1}=\Gamma$, and moreover that it is the union of conjugacy classes, so that it is invariant under conjugation, $g \Gamma g^{-1}=\Gamma$ for any $g$ in $G$ \cite{spectralgraphtheory}.
We choose $\Gamma$ not to include the identity element and we note that the choice of $\Gamma$ is not unique.
The Cayley graph has the group elements as vertices, and we place an edge between $g \in G$ and $h \in G$ if $h g^{-1} \in \Gamma$.
The result is a simple undirected graph.
Examples of Cayley graphs for the groups $\Z_5$ and $D_4$ are shown in Fig.~\ref{fig:examples of Cayley graphs}.
Given any graph, its Laplacian is defined as \cite{spectralgraphtheory} 
\begin{equation}
    \Delta = D-A \ ,
\end{equation}
where $A$ is the adjacency matrix and $D$ is the degree matrix.
Each of these matrices acts on the vector space of graph vertices, which in the case of a Cayley graph can be identified with the group algebra $\C[G]$.
The degree matrix is always diagonal, and in this case $D=\abs{\Gamma} \mathbbm{1}$.
The adjacency matrix $A$ is given by
\begin{equation}
    A_{gh} = \begin{cases}1 & g h^{-1} \in \Gamma\\ 0 & \mathrm{otherwise}\end{cases}
\end{equation}
for group elements $g,h$.
On a basis element, one has
\begin{equation}
    A \ket{g} \equiv \sum_h A_{hg} \ket{h} = \sum_{k \in \Gamma} \ket{gk} = \sum_{k \in \Gamma} \ket{gk^{-1}} = \sum_{k \in \Gamma} R_k \ket{g} \ ,
\end{equation}
where $R_k$ is the right regular representation, and we used the closure of $\Gamma$ under inversion.
Therefore as an operator on $\C[G]$, 
\begin{equation}
    A = \sum_{k \in \Gamma} R_k = \bigoplus_j \mathbbm{1}_j \otimes \pqty{\sum_{k \in \Gamma} \rho_j(k)} \ ,
\end{equation}
where we used the Peter-Weyl decomposition of $R_k$ \eqref{eq:translations peter weyl}.
Then we see that
\begin{equation}
    \pqty{\sum_{k \in \Gamma} \rho_j(k)} \rho_j(g) = \sum_{k \in \Gamma} \rho_j(kg) = \sum_{k \in \Gamma} \rho_j(g k g^{-1} g) = \rho_j(g) \pqty{\sum_{k \in \Gamma} \rho_j(k)} \ ,
\end{equation}
where we used the closure of $\Gamma$ under conjugation.

Hence the operator $\pqty{\sum_{k \in \Gamma} \rho_j(k)}$ commutes with the irreducible representation $\rho_j$ and as such is proportional to the identity by Schur's lemma \cite{Serre}.
The constant of proportionality can be readily computed by taking a trace.
This therefore implies 
\begin{equation}
    A = \sum_j \lambda_j P_j \ , \quad \quad \lambda_j = \frac{1}{\dim{j}} \sum_{k \in \Gamma} \chi_j(k) \ ,
\end{equation}
where $P_j = \sum_{mn} \ket{jmn}\bra{jmn}$ is the projector onto the $j$-th representation subspace, and $\chi_j$ is the character of the irrep labelled $j$.
Therefore the Laplacian of the Cayley graph is given by
\begin{equation}\label{eq:laplacian finite group}
    \Delta = \sum_j f(j) P_j \ , \quad \quad f(j)= \abs{\Gamma}-\frac{1}{\mathrm{dim}(j)} \sum_{k \in \Gamma} \chi_j(k) \ ,
\end{equation}
which is the same form as the electric Hamiltonian in the representation basis, \eqref{eq:electric hamiltonian rep basis}. For any finite group, this formula defines the electric energy $f(j)$ to be assigned to each irrep.

\medskip

We give some examples of this construction.
For the group $\Z_N$ it is natural to construct the electric eigenvalues $f(j)$ with the generating set $\Gamma = \{\xi, \xi^{-1}\}$ where $\xi$ is a generator of $\Z_N$, which results in
\begin{equation}
    \label{eq:fj ZN}
    f(j)=f(N-j) = 4\sin^2\pqty{\frac{\pi j}{N}} \ ,
\end{equation}
which is the same as in \cite{Ercoetal1}.
Moreover for large $N$,
\begin{equation}
    f(j) \to \frac{4\pi^2}{N^2} j^2 \qquad N \mathrm{~large} \ ,
\end{equation}
which is proportional to the Casimir eigenvalues of $\U(1)$ gauge theory \cite{Ercoetal2}.
Thus both a truncation of $\U(1)$ theory and proper $\Z_N$ theory naively approach $\U(1)$ theory for large $N$, albeit in different ways.
One can however choose a different generating set, such as $\Gamma = \{\xi, \xi^{-1}, \xi^2, \xi^{-2}\}$ and the corresponding eigenvalues would be
\begin{equation}
    f(j)=f(N-j) = 4\sin^2\pqty{\frac{\pi j}{N}}+4\sin^2\pqty{\frac{2\pi j}{N}} \ .
\end{equation}
For the dihedral group $D_4$ we can choose for example 
\begin{equation*}
    \Gamma = \Gamma_1 = \{r,r^3,s,r^2s\},
\end{equation*}
which gives rise to the eigenvalues $f(j)$  shown in Table \ref{tab:fval}, where the representations are ordered like in the character table in Table \ref{tab:char} in Appendix \ref{sec:DN groups}.
Note that $\Gamma_1$ generates the whole group.

\begin{table}[t]
    \centering
    \begin{tabular}{lcccccc}
        \toprule
         & & \multicolumn{5}{c}{$f(j)$} \\
        \cmidrule(l){3-7}
        \hspace{5em}$\Gamma$ & $j$ & 0 & 1 & 2 & 3 & 4\\
        \midrule
        $\Gamma_1 = \{r, r^3, s, r^2 s \}$
                 & & 0 & 4 & 4 & 8 & 4 \\[5pt]
        $\Gamma_2 = \{r, r^3, s, r s, r^2 s, r^3 s \}$
                       & & 0 & 8 & 8 & 8 & 6 \\[5pt]
        $\Gamma_3 = \{r, r^2, r^3 \}$
                 & & 0 & 4 & 0 & 4 & 5 \\
        \bottomrule
    \end{tabular}
    \caption{Eigenvalues of the single-link electric Hamiltonian $f(j)$ for the finite group $D_4$, with three choices of generating sets: $\Gamma_1$, $\Gamma_2$, and $\Gamma_3$ respectively.}
    \label{tab:fval}
\end{table}

By looking at its character table, we may in fact classify all possible choices of $\Gamma$ for $D_4$.
In fact, $D_4$ has five conjugacy classes:
\begin{equation*}
    C_0 = \{e\}, \quad
    C_1 = \{r, r^3\}, \quad
    C_2 = \{r^2\}, \quad
    C_3 = \{s, r^2s\}, \quad
    C_4 = \{rs, r^3s\}.
\end{equation*}
One can check that, as is generally true, $\sum_i \abs{C_i}=8=\abs{G}$.
In this case, all conjugacy classes are invariant under inversion, i.e.
$C_i^{-1}=C_i$.
 Hence any union of the $C_i$'s, $i \neq 0$ is a valid choice for $\Gamma$.
There are $2^4$ such possibilities.
Note that this is not true in general, in which case one must choose conjugacy classes to ensure that $\Gamma^{-1}=\Gamma$.
In the next sections we will consider in more detail two specific cases:
\begin{equation*}
        \Gamma_2 = C_1 \cup C_3 \cup C_4=\{r, r^3, s, rs, r^2s, r^3s\}
        \quad \text{and} \quad
        \Gamma_3 = C_1 \cup C_2 = \{r, r^2, r^3\}.
\end{equation*}
The choice of $\Gamma_2$ is especially interesting, because it corresponds to the Hamiltonian arising from the transfer-matrix procedure when $h_B$ is the real part of the trace of the faithful irrep of $D_4$; therefore, this choice gives rise to a manifestly Lorentz-invariant theory.
Note that also $\Gamma_2$ generates the whole group.
On the other hand, the set $\Gamma_3 = \{r, r^2, r^3\}$ does not generate the whole group, but only the subgroup of rotations;
this is reflected in the electric eigenvalues in Table \ref{tab:fval}, with the electric ground state being two-fold degenerate on each link.

\subsubsection{Action formulation and Lorentz invariance}\label{sec:action formulation}

The Kogut-Susskind Hamiltonian \eqref{eq:kog suss hE hB} may be obtained via the transfer-matrix formulation from the Euclidean Wilson action \cite{CreutzTransferMatrix,KogRev}
\begin{equation}
    \label{eq:wilson action}
    S = -\frac{2}{g^2} \sum_{\square} \Re \tr{\rho(g_\square)} \ ,
\end{equation}
where $g$ is the coupling.
In the path-integral formulation, the lattice is fully discretized and thus plaquettes extend also in the time direction.
The action \eqref{eq:wilson action} is also perfectly valid for finite groups, as one simply replaces the integration measure over the Lie group with a sum over the elements of a finite group. The representation $\rho$ can be chosen to be any faithful representation of the finite group (not necessarily irreducible).
One may then repeat the transfer-matrix formulation for an arbitrary finite group \cite{TransferMatrixFiniteGroup}.
Starting from the action \eqref{eq:wilson action}, the transfer-matrix procedure gives rise to a Hamiltonian of the form \eqref{eq:generalized ym hamiltonian} that we've described, with
\begin{equation}
    \label{eq:lorentz invariance relations}
    \Gamma = \{g \in G,\,\, g \neq 1, \mathrm{max} \bqty{\Re \tr{\rho(g)}}  \} \quad \quad \quad h_B=-2\Re\tr\rho \ .
\end{equation}
In other words, the magnetic Hamiltonian is directly inherited from the action, while the electric Hamiltonian takes the form of the finite-group Laplacian with a specific choice of the generating set $\Gamma$ for the Cayley graph. In the example of the gauge group $D_4$, if we choose the faithful, two-dimensional irrep for $h_B$, then $\Re\tr\rho(g)$ can equal $2, 0, -2$ on the different conjugacy classes (see the character table of $D_4$ in Table \ref{tab:char}). Since $\Re\tr\rho(1)=2$, then $\Gamma$ consists of all group elements $g$ such that $\Re\tr\rho(g)=0$. This gives rise to the generating set $\Gamma_2$ anticipated in Section \ref{sec:finitegrouplaplacian}.

These considerations are especially important for the Lorentz invariance of the theory. While the lattice discretization breaks the Lorentz symmetry to the subgroup of Euclidean cubic rotations, as long as this subgroup is preserved one expects to recover Lorentz invariance in the continuum limit. In particular, the action \eqref{eq:wilson action} is manifestly invariant under Euclidean cubic rotations and, therefore, one expects that it gives rise to a  Hamiltonian which describes a Lorentz-invariant theory in the continuum. Intuitively, a Lorentz transformation can swap the electric and magnetic fields, and it is therefore not surprising that in a Lorentz-invariant theory the electric and magnetic Hamiltonians must satisfy specific relations with each other.

In particular, finite-group Hamiltonians of the form \eqref{eq:generalized ym hamiltonian} which however do not respect the relations \eqref{eq:lorentz invariance relations} cannot arise from an action of the form \eqref{eq:wilson action}. For example, they could come from an action in which plaquettes extending in one direction (the \say{time} direction) are weighted differently. For such Hamiltonians, it is unclear whether they describe a Lorentz-invariant theory. This includes in particular setting $h_E(j)=j^2$ for subgroups of $\U(1)$ and $h_E(j)=j(j+1)$ for subgroups of $\SU(2)$ in \eqref{eq:electric hamiltonian rep basis}, while keeping $h_B$ unchanged. In all such cases, the remnant Lorentz symmetry is explicitly broken. While Lorentz symmetry is required in particle physics applications, it might not be necessarily required in other cases, such as some condensed matter applications, and one may thus independently choose $\Gamma$ and $h_B$.

\subsubsection{Classification of the possible theories, and other models}\label{sec:classification}

The construction of finite group Yang-Mills gauge theories with Hamiltonian \eqref{eq:generalized ym hamiltonian} involves a few arbitrary choices which can be classified.
Since the Hilbert space is fixed to the physical, gauge-invariant Hilbert space $\mathcal{H}_{\mathrm{phys}}$, the only possible choices involve the various terms in the Hamiltonian.
Given a gauge group $G$ in $d$ spatial dimensions, one may arbitrarily choose:
\begin{enumerate}
    \item A set $\Gamma$ of group elements which does not contain the identity, and is invariant under both inversion and conjugation $\Gamma^{-1}=\Gamma$ and $g \Gamma g^{-1}$.
    \item A choice for the magnetic Hamiltonian $h_B = h_B(g_\square)$.
Since it must be real and satisfy $h_B(g_1 g_\square g_1^{-1})= h_B(g_\square)$, i.e.
it is a \textit{class function}, by a well-known result \cite{Serre} it may be expanded as a sum of characters of irreducible representations, $h_B(g) = \sum_j c_j \chi_j(g)$ for coefficients $c_j$ which may be chosen arbitrarily, while ensuring that $h_B(g)$ is real. Most typically $h_B = -2 \Re\tr\rho$, where $\rho$ is some faithful representation (not necessarily an irrep).
    \item If present, possible choices of representations and Hamiltonians in the matter sector.
\end{enumerate}
Further considerations involve the Lorentz symmetry, as explained in Section \ref{sec:action formulation}.
In most applications, one might want to choose faithful representations of the gauge group; otherwise one is effectively choosing a different, smaller gauge group.

We note that the above construction allows further generalizations.
In particular, the discretized $d$-dimensional space does not have to take the form of a hypercubic lattice, but more generally can be a Bravais or non-Bravais lattice, or even a cell complex.
No difference arises for the electric term, which is link-based, and the plaquette variable in the magnetic term is replaced by an an elementary closed loop in the lattice.
Moreover, as we will see in Section \ref{sec:physical Hilbert space}, the description of the gauge-invariant Hilbert space in terms of spin-network states is valid generally on an arbitrary graph discretizing spacetime.

\section{The physical Hilbert space}\label{sec:physical Hilbert space}

As we remarked in Section \ref{sec:basic hilbert space}, while the overall Hilbert space of the pure gauge theory is $\mathcal{H}_{\mathrm{tot}} = \bigotimes_{\mathrm{links}} \C[G]$, only those states that satisfy the so-called \say{Gauss' law} constraint \eqref{eq:lattice gauss law} are to be considered physical \cite{KogSuss, Osborne, Tong}.
For gauge theories based on most compact Lie groups, the Wilson loops (despite being overcomplete) span the space of gauge-invariant states \cite{Sengupta, Durhuus}.
This, however, is not necessarily true for finite groups \cite{Sengupta, Cui}; this means that in some cases, it is possible to construct different gauge-invariant states, which nevertheless have identical Wilson loops.
We mention that this difficulty does not arise for Abelian finite groups such as $\Z_N$, in which case the Wilson loops \textit{do} span the physical Hilbert space $\mathcal{H}_{\mathrm{phys}}$.
As will be discussed in Section \ref{sec:spin networks pure gauge}, the gauge-invariant Hilbert space for pure gauge theories may be described in terms of \textit{spin network states}. This basis turns out to be particularly suitable for finite gauge groups and in Section \ref{sec:spin networks dimension} we give a simple formula to compute the dimension of the physical Hilbert space for any finite gauge group.

\subsection{Spin network states}\label{sec:spin networks pure gauge}

The physical Hilbert space of pure gauge theories with either Lie or finite gauge group can be explicitly described in terms of \textit{spin network states} \cite{Baez, Burgio}.
Spin-network states can be defined indifferently when the $d$-dimensional space is discretized as an arbitrary graph, and are thus valid in arbitrary dimension with arbitrary lattices and boundary conditions.

\medskip

The first key observation is that the action of the Gauss' law operator \eqref{eq:lattice gauss law} is block-diagonal in the representation basis, as can be seen from \eqref{eq:translations peter weyl}.
Then starting from the Hilbert space in the representation basis \eqref{eq:peterweyl}, we can, as usual, permute the summation and product, obtaining
\begin{equation}
    \label{eq:hilbert space permuted}
    \mathcal{H}_{\text{tot}} = \bigotimes_{\mathrm{links}} \bigoplus_{j \in \Sigma} V_j^* \otimes V_j=\bigoplus_{\{j\} \in \{\Sigma\}} \bigotimes_{l \in \mathrm{links}}  V_{j_l}^* \otimes V_{j_l} \ ,
\end{equation}
where now $\{j\}$ is an assignment of an irrep $j_l$ to each lattice link $l$, and $\{\Sigma\}$ is the set of the possible assignments.
The second key observation is that the gauge transformations \eqref{eq:lattice gauss law} are given by an independent group-valued variable $g_x$ at each site $x$ of the lattice.

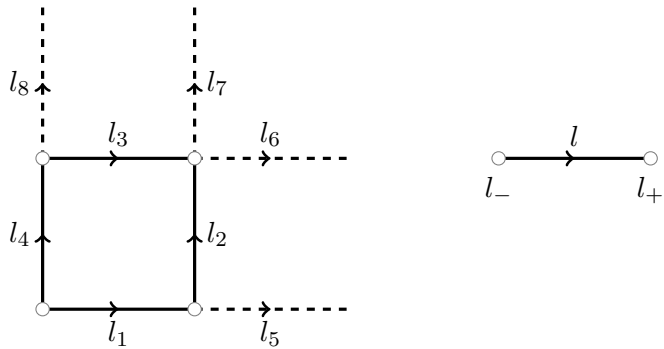
\begin{figure}[t]
    \centering
    \begin{tikzpicture}[
    font=\small,
    site/.style={circle, inner sep=0pt, minimum size=5pt, draw=gray, fill=white},
    inside link/.style={black, very thick, ->-=0.5},
    outside link/.style={black, very thick, dashed, ->-=0.5},
    testo/.style={midway, black}
]
\draw[inside link]  (0,0) -- (2,0) node[testo, below] {$l_1$};
\draw[inside link]  (2,0) -- (2,2) node[testo, right] {$l_2$};
\draw[inside link]  (0,2) -- (2,2) node[testo, above] {$l_3$};
\draw[inside link]  (0,0) -- (0,2) node[testo, left ] {$l_4$};
\draw[outside link] (0,2) -- (0,4) node[testo, left ] {$l_8$};
\draw[outside link] (2,2) -- (2,4) node[testo, right] {$l_7$};
\draw[outside link] (2,2) -- (4,2) node[testo, above] {$l_6$};
\draw[outside link] (2,0) -- (4,0) node[testo, below] {$l_5$};
\foreach \x in {0, 2} \foreach \y in {0, 2} \node[site] at (\x, \y) {};

\draw[inside link] (6, 2) -- (8, 2) node [testo, above] {$l$};
\node[site, label=below:{$l_{-}$}] at (6, 2) {};
\node[site, label=below:{$l_{+}$}] at (8, 2) {};
\end{tikzpicture}
    \caption{\emph{Left}: a $2\times 2$ square lattice with periodic boundary conditions, showing the labels of the links. 
        \emph{Right}: labelling of sites attached to a link.}
    \label{fig:periodic plaquette}
\end{figure}

Moreover, due to \eqref{eq:translations peter weyl} the gauge transformation associated to one site $x$ acts at most on one of the spaces $V_j$ or $V_j^*$ associated to a link, but it cannot act on both.
One can then split the two vector spaces $V_j$ and $V_j^*$ associated with each link and reorder the $V$'s in the tensor product over links so that $V_j$'s are now grouped together according to the \textit{sites} and not the links,
\begin{equation}
    \mathcal{H}_{\text{tot}} = \bigoplus_{\{j\} \in \{\Sigma\}} \bigotimes_{x \in \mathrm{sites}} \pqty{\bigotimes_{l_-=x} V_{j_l}^*} \otimes \pqty{\bigotimes_{l_+=x} V_{j_l} }\ ,
\end{equation}
where by $l_+$ and $l_-$ we denote respectively the target and source vertex of link $l$ (see Fig.~\ref{fig:periodic plaquette}).

\medskip

We can repeat the same set of operations for the gauge transformation operator \eqref{eq:gauss law operator}, which is therefore given by
\begin{equation}
    \mathcal{G}(\{g_x\}) = \bigoplus_{\{j\} \in \{\Sigma\}} \bigotimes_{x \in \mathrm{sites}} \pqty{\bigotimes_{l_-=x} \rho_{j_l}^*(g_x)} \otimes \pqty{\bigotimes_{l_+=x} \rho_{j_l}(g_x) }\ .
\end{equation}
In the above decomposition, the gauge transformations now act independently for each $x$ and the Gauss' law constraint \eqref{eq:lattice gauss law} gives the physical Hilbert space
\begin{equation}
    \label{eq:spin network Hilbert space}
    \mathcal{H}_{\mathrm{phys}} = \bigoplus_{\{j\} \in \{\Sigma\}} \bigotimes_{x \in \mathrm{sites}} \mathrm{Inv}\bqty{\pqty{\bigotimes_{l_-=x} V_{j_l}^*} \otimes \pqty{\bigotimes_{l_+=x} V_{j_l} }} \ .
\end{equation}
Given a representation $\rho$ (not necessarily irreducible) with representation space $V_\rho$, the set of invariant vectors $\mathrm{Inv}(V_\rho)$ is the set of vectors $v \in V_\rho$ such that $\rho(g) v = v$ for all $g\in G$.
Note that this is a separate notion from that of an \say{invariant subspace}.
The characterization of the Hilbert space \eqref{eq:spin network Hilbert space} implies that any physical, gauge-invariant state $\ket{\Psi}$ (i.e.
a state which satisfies the Gauss' law \eqref{eq:lattice gauss law}) may be expanded in a basis of \textit{spin network states},
\begin{equation}
    \ket{\Psi} = \sum_{ \{j\} } \sum_{ A } \Psi(\{j\}; A) \ket{\{j\},A}, \qquad \ket{\{j\},A} = \bigotimes_{x \in \mathrm{sites}} \ket{\{j\}_x,a_x},
\end{equation}
where $\{j\}$ is an assignment of irreps to lattice links and $A=(a_1, \ldots a_V)$ is a multi-index which labels the choice of a basis element of invariant states at each site.
With $\{j\}_x$ we denote the irreps assigned to the links connected to site $x$.

For a hypercubic lattice in $d$ dimensions with periodic boundary conditions, each site is connected to $2d$ links and therefore $2d$ terms appear in the tensor product within each $\mathrm{Inv}$ in \eqref{eq:spin network Hilbert space}.
If instead we choose open boundary conditions, the sites in the bulk will again be connected to $2d$ links, but the sites on the boundary will be connected to fewer links and thus fewer terms will appear in the tensor product for those sites.
In the general case, the number of terms in the tensor product within each $\mathrm{Inv}$ will thus depend on the site.
We choose to work directly with the spaces of invariant vectors rather than with spaces of intertwiners more commonly employed in the literature on spin-network states \cite{Baez, Burgio}.
We also would like to note that the physical Hilbert space \eqref{eq:spin network Hilbert space} contains \textit{all} gauge-invariant states, possibly also including states in sectors with a non-contractible Wilson line.

\medskip

The calculation of a basis of invariant states (or, equivalently, of the intertwiners) can be difficult in the Lie group case, especially since they admit infinitely many irreps.
On the other hand, since the number of links connected to each site is finite and independent of the lattice volume, one needs only compute the invariant states of a finite number of tensor product representations which does not scale with the lattice volume.

This can be achieved in practice by explicitly writing out the matrices of the tensor product representation $\rho(g) \equiv \pqty{\bigotimes_{l_-=x} \rho_{j_l}^*} \otimes \pqty{\bigotimes_{l_+=x} \rho_{j_l} }$ and solving the simultaneous equations $\rho(g) v = v$ for a set of generators of $G$.
In a $d$-dimensional periodic hypercubic lattice, the number of terms in the tensor product equals $2d$ and the maximum dimension of the tensor product representation is bounded by $\pqty{\dim{j}}^{2d} \leq \abs{G}^d$, owing to $\sum_j \pqty{\dim{j}}^2 = \abs{G}$, independently from the lattice volume.

As an example, we work out explicitly the case of a $2\times 2$ square lattice with periodic boundary conditions.
As shown in Fig.~\ref{fig:periodic plaquette}, this system has four vertices and eight links.
Expanding explicitly \eqref{eq:spin network Hilbert space} we see that in this case
\begin{align}
    \mathcal{H}_{\mathrm{phys}} &= \bigoplus_{j_1, \ldots j_8}
    \mathrm{Inv}\bqty{V_{j_1}^* \otimes V_{j_4}^* \otimes V_{j_5} \otimes V_{j_8}} \otimes
    \mathrm{Inv}\bqty{V_{j_5}^* \otimes V_{j_2}^* \otimes V_{j_1} \otimes V_{j_7}}  \otimes \\
    &\quad\otimes\mathrm{Inv}\bqty{V_{j_6}^* \otimes V_{j_7}^* \otimes V_{j_3} \otimes V_{j_2}} \otimes
    \mathrm{Inv}\bqty{V_{j_3}^* \otimes V_{j_8}^* \otimes V_{j_6} \otimes V_{j_4}} \ . \nonumber
\end{align}
Now consider a single invariant subspace $\mathrm{Inv}\bqty{V_{j_1}^* \otimes V_{j_2}^* \otimes V_{j_3} \otimes V_{j_4}}$ with arbitrary assignment of irreps.
This vector space admits an orthonormal basis $\{\ket{j_1 j_2 j_3 j_4 ; a}\}$ where $1 \leq a \leq \dim\mathrm{Inv}\bqty{V_{j_1}^* \otimes V_{j_2}^* \otimes V_{j_3} \otimes V_{j_4}}$ indexes the basis vector.
We can expand the basis vectors explicitly in terms of the bases of the $V_j$ as (see also the discussion around \eqref{eq:change of basis})
\begin{equation}
    \label{eq:invariant states at a site}
    \ket{j_1 j_2 j_3 j_4 ; a} = \sum_{m_1, m_2, n_3, n_4} \psi\pqty{j_1 m_1 j_2 m_2 j_3 n_3 j_4 n_4 ; a} \ket{j_1 m_1} \otimes \ket{j_2 m_2} \otimes \ket{j_3 n_3} \otimes \ket{j_4 n_4} \ .
\end{equation}
The basis vectors can be chosen to be orthonormal.
By virtue of spanning the space $\mathrm{Inv}\bqty{V_{j_1}^* \otimes V_{j_2}^* \otimes V_{j_3} \otimes V_{j_4}}$, they are invariant vectors of the tensor product representation $\rho \equiv \rho_{j_1}^* \otimes \rho_{j_2}^* \otimes \rho_{j_3} \otimes \rho_{j_4}$; as such, they satisfy $\rho(g) \ket{j_1 j_2 j_3 j_4 ; a} = \ket{j_1 j_2 j_3 j_4 ; a}$ for all $g \in G$.
The coefficients of the expansion $\psi\pqty{j_1 m_1 j_2 m_2 j_3 n_3 j_4 n_4 ; a}$ may be easily computed, for example by writing the tensor product representation matrices $\rho(g)$ explicitly and then solving the simultaneous equations $\rho(g) v = v$.
The dimension of the space of invariant vectors depends on the four representations assigned to the relevant site.
Now let $A = \pqty{a_1, a_2, a_3, a_4}$, which implicitly depends on $\{j\}$ (because the range of each $a_x$ depends on the irreps assigned to site $x$).
Given any assignment of irreps $\{j\}$, $A$ is a choice of a basis vector of invariant states at the four sites.
Therefore an orthonormal basis for the gauge invariant Hilbert space is given by
\begin{equation}
    \ket{ \{j\};A} = \ket{j_1 j_4 j_5 j_8 ; a_1} \otimes \ket{j_5 j_2 j_1 j_7 ; a_2} \otimes \ket{j_6 j_7 j_3 j_2 ; a_3} \otimes \ket{j_3 j_8 j_6 j_4 ; a_4} \ ,
\end{equation}
for any possible assignment $\{j\}$ of irreps to links, and then all possible choices $A$ of an invariant vector at each of the four sites. The spin-network states $\ket{ \{j\};A}$ then form a basis of the gauge-invariant Hilbert space $\mathcal{H}_{\mathrm{phys}}$.
Expanding the tensor product, we find an explicit expression for these states in terms of the representation basis,
\begin{align}
    \label{eq:spin network states 2x2}
    \ket{ \{j\};A} &= \sum_{n_1, \ldots n_8} \sum_{m_1, \ldots m_8} \psi\pqty{j_1 m_1 j_4 m_4 j_5 n_5 j_8 n_8 | a_1} \psi\pqty{j_5 m_5 j_2 m_2 j_1 n_1 j_7 n_7 | a_2} \times \nonumber \\
    &\quad \quad \times \psi\pqty{j_6 m_6 j_7 m_7 j_3 n_3 j_2 n_2 | a_3} \psi\pqty{j_3 m_3 j_8 m_8 j_6 n_6 j_4 n_4 | a_4} \times \\
    &\quad \quad \times \ket{j_1 m_1 n_1} \otimes \ket{j_2 m_2 n_2} \otimes \cdots \ket{j_8 m_8 n_8} \nonumber \ ,
\end{align}
where we have restored the ordering of the vector spaces $V_j$s and used again the shorthand $\ket{jmn}=\ket{jm}\otimes \ket{jn}$. We note in particular that despite having introduced a splitting of the variables at each link, in the final answer this splitting disappears and the spin-network states can be entirely expressed in terms of the representation basis $\ket{jmn}$.

\subsection{The dimension of the physical Hilbert space}\label{sec:spin networks dimension}

As we have seen in the previous section, spin network states give an explicit description of the physical Hilbert space $\mathcal{H}_{\mathrm{phys}}$ as
\begin{equation}
    \mathcal{H}_{\mathrm{phys}} =\bigoplus_{\{\rho\} \in \{\Sigma\}} \bigotimes_{v \in \mathrm{sites}} \mathrm{Inv}\bqty{\pqty{\bigotimes_{l_- = v}  V_{\rho_l}^*} \otimes \pqty{\bigotimes_{l_+ = v}  V_{\rho_l}}} \ ,
\end{equation}
where $\mathrm{Inv}(\rho)$ is the space of invariant vectors of the representation $\rho$, $\{\rho\}$ is an assignment of irreps to links and $\{\Sigma\}$ is the set of such possible assignments. For a finite group,
\begin{equation}
    \label{eq:invariant vectors counting}
    \dim{\mathrm{Inv}(\rho)} = \frac{1}{\abs{G}} \sum_{g \in G} \chi_\rho(g) \ ,
\end{equation}
where $\chi_\rho$ is the character of $\rho$. A proof of this result can be found in Appendix \ref{sec:counting invariant states}. This fact can be used to obtain a general formula for the dimension of the $\mathcal{H}_{\mathrm{phys}}$, which is valid for any lattice in any dimension with any boundary conditions. On a connected lattice with $L$ links and $V$ sites, we will show that
\begin{equation}
    \label{eq:physical hilbert space dimension}
    \dim{\mathcal{H}_{\mathrm{phys}}} = \sum_C \pqty{\frac{\abs{G}}{\abs{C}}}^{L-V} \ ,
\end{equation}
where the sum runs over all conjugacy classes $C$ of the group, and $\abs{C}$ is the size of $C$. The ratio $\abs{G}/\abs{C}$ is always an integer by the orbit-stabilizer theorem \cite{Serre}. 
Since for a connected graph $L-V \geq -1$, the $\dim{\mathcal{H}_{\mathrm{phys}}}$ in \eqref{eq:physical hilbert space dimension} is always an integer.
This is clear for $L-V \geq 0$; when $L-V=-1$ the graph is a tree and since $\sum_C \abs{C}=\abs{G}$, we find $\dim{\mathcal{H}_{\mathrm{phys}}}=1$; this is to be expected because, on a tree, all the physical degrees of freedom can be rotated away by gauge transformations.

Using \eqref{eq:invariant vectors counting}, together with the fact that the character of a tensor product is given by the product of the characters, we may readily prove \eqref{eq:physical hilbert space dimension}. From the general formula for the gauge-invariant Hilbert space, we have
\begin{align}
    \dim{\mathcal{H}_{\mathrm{phys}}} &=\sum_{j_1 j_2 \cdots j_L} \prod_{x \in \mathrm{sites}} \dim\mathrm{Inv}\bqty{\pqty{\bigotimes_{l_- = x}  V_{\rho_l}^*} \otimes \pqty{\bigotimes_{l_+ = x}  V_{\rho_l}}} = \nonumber \\
    &=\frac{1}{\abs{G}^V}\sum_{j_1 j_2 \cdots j_L} \sum_{g_1 g_2 \cdots g_V} \prod_{x \in \mathrm{sites}} \pqty{\prod_{l_- = x}  \chi_{j_l}^*(g_x)}  \pqty{\prod_{l_+ = x} \chi_{j_l}(g_x)} \nonumber \ .
\end{align}
Within the product over all sites, there are exactly $2L$ factors of characters $\chi$, as each link contributes two representation spaces $V$ and each representation space gives rise to a character. Thus grouping characters by link, we obtain
\begin{equation}
    \dim{\mathcal{H}_{\mathrm{phys}}} =\frac{1}{\abs{G}^V} \sum_{g_1 g_2 \cdots g_V} \prod_{l = \expval{xx'} \in \mathrm{links}} \expval{g_x, g_{x'}} \ .
\end{equation}
where we denoted $\expval{g, h} = \sum_j \chi_j(g)^* \chi_j(h)$. It is a well-known result that $\expval{g, h}$ is zero unless $g$ and $h$ belong to the same conjugacy class, in which case $\expval{g, h} = \abs{G} / \abs{C}$ where $C$ is the conjugacy class of both $g$ and $h$ \cite{Serre}. If any two adjacent sites $x$ and $x'$ have $g_x$ and $g_{x'}$ in different conjugacy classes, then $\expval{g_x, g_{x'}}=0$ and the corresponding term in the sum is zero. Assuming that the lattice is connected, this implies that the product over all links is zero unless all the $g_x$ at each site $x$ belong to the same conjugacy class. Then, since $\expval{g_x, g_{x'}}$ is constant on conjugacy classes, we can write
\begin{equation}
    \dim{\mathcal{H}_{\mathrm{phys}}}
    =\frac{1}{\abs{G}^V} \sum_C \sum_{g_1 g_2 \cdots g_V \in C} \frac{\abs{G}^L}{\abs{C}^L} = \sum_C  \pqty{\frac{\abs{G}}{\abs{C}}}^{L-V} \ ,
\end{equation}
which concludes the proof.
In the Abelian case the above formula simplifies as all conjugacy classes are singlets and therefore $\dim{\mathcal{H}_{\mathrm{phys}}} = \abs{G}^{L-V+1}$.
Thus finite Abelian groups have the largest physical Hilbert space among all groups of the same order. 
For periodic boundary conditions in a hypercubic lattice, $L=Vd$ and as such $\dim{\mathcal{H}}=\abs{G}^{Vd}$, while $\dim{\mathcal{H}_{\mathrm{phys}}} \approx \abs{G}^{V (d-1)}$, so that the physical Hilbert space has roughly the same size as the overall Hilbert space in one lower dimension. 
Nonetheless, both spaces grow exponentially with the lattice size.

\medskip

As a further example, we consider the dimension of the Hilbert space for pure $D_4$ gauge theory.
Using \eqref{eq:physical hilbert space dimension}, we find for $G=D_4$ on a lattice with $L$ links and $V$ sites (see also \cite{marchesethesis}), 
\begin{equation}
    \dim{\mathcal{H}_{\mathrm{phys}}} = 8^{L-V}\pqty{2+\frac{3}{2^{L-V}}} \ .
\end{equation}
The dimension of the physical Hilbert space for some two-dimensional finite square lattices in $2+1$ dimensions is shown in Table \ref{tab:numstates}. We see that its size grows quickly with the lattice size. We point out that even for a $2 \times 2$ periodic lattice with a small group such as $D_4$ it is not practical to write down all possible gauge-invariant states. Unless the structure happens to be very sparse, writing down the $8960$ physical basis elements in terms of the $\abs{G}^L = 8^8$ basis elements in the representation basis using $4 \mathrm{B}$ floating point numbers would require roughly $600 \,\mathrm{GB}$ of memory. For a  $3 \times 3$ periodic lattice this number rises to $20\, \mathrm{YB}$ or $2 \cdot 10^{16}\, \mathrm{GB}$.

Finally, we remark that since matter fields are site-based, the spin-network states may be extended to this case as well; the physical Hilbert space would then be given again by \eqref{eq:spin network Hilbert space} with an extra factor of the matter Hilbert space at each site within each $\mathrm{Inv}$. The detailed description of the gauge-invariant Hilbert space with matter fields will be given in a future publication.

\begin{table}[t]
    \centering
    \begin{tabular}{clrrcr}
        \toprule
        Size~~~ & BCs & $L$ & $V$ & $L-V$ &$\dim{\mathcal{H}_{\mathrm{phys}}}$\\
        \midrule
        \multirow{2}{3em}{$2 \times 2$}
            & open & $4$ & $4$ & $0$ & $5$\\
            & periodic & $8$ & $4$ & $4$ & $8960$ \\[5pt]
        \multirow{2}{3em}{$2 \times 3$}
            & open & $7$ & $6$ & $1$ & $28$ \\
            & periodic & $12$ & $6$ & $6$ & $536576$ \\[5pt]
        \multirow{2}{3em}{$3 \times 3$}
            & open & $12$ & $9$ & $3$ & $1216$ \\
            & periodic & $18$ & $9$ & $9$ & $269221888$ \\
        \bottomrule
    \end{tabular}
    \caption{Dimension of the physical subspace of $D_4$ gauge theory on some small lattices in $2+1$ dimensions. $L$ is the number of links and $V$ is the number of vertices.}
    \label{tab:numstates}
\end{table}

\section{Dihedral gauge theory on a small periodic lattice}

In this section we consider pure gauge theory with gauge group $G=D_4$, the dihedral group with eight elements, on a small $2 \times 2 $ periodic lattice (see Fig.~\ref{fig:periodic plaquette}). We compute the Hamiltonian in the gauge-invariant spin-network basis and diagonalize it exactly. As remarked in Section \ref{sec:spin networks dimension}, the physical Hilbert space of this theory has dimension equal to $8960$ and it's not practical to store the gauge-invariant states directly. Instead, we first compute numerically the basis of invariant states at a site for all possible combinations of irreps assigned to the four links attached to the site (i.e. we compute the coefficients $\psi$ of \eqref{eq:invariant states at a site}). Using these coefficients we then compute the matrix elements of the electric and magnetic Hamiltonians separately in the spin-network basis \eqref{eq:spin network states 2x2}. The electric Hamiltonian is diagonal, and the magnetic Hamiltonian is off-diagonal. In units of $\lambda_E + \lambda_B$ the Hamiltonian can be written as $H = \lambda H_E+(1-\lambda) H_B$ for $\lambda \in [0,1]$, where $\lambda = \lambda_E/(\lambda_E+\lambda_B)$. In practice, for each $\lambda$, $H$ is a $8960 \times 8960$ matrix. As expected for spin-network states \cite{Burgio}, we find $H$ to be very sparse: less than $1\%$ of the elements are non-zero.

The electric and magnetic Hamiltonians were chosen as in \eqref{eq:generalized ym hamiltonian}. In particular, we chose $h_B = -2 \tr \rho_4$ where $\rho_4$ is the two-dimensional irrep of $D_4$ and considered the three different choices of the set $\Gamma$ for $h_E$ described in Section \ref{sec:finitegrouplaplacian}. These are $\Gamma_1 = \{r,r^3,s,r^2s\}$, $\Gamma_2=\{r, r^3, s, rs, r^2s, r^3s\}$ and $\Gamma_3 = \{r, r^2, r^3\}$. We recall that the electric Hamiltonian is two-fold degenerate on each link with the choice of $\Gamma_3$ but is not degenerate with $\Gamma_1$ or $ \Gamma_2$. The choice of $\Gamma_2$, unlike the other two, gives rise to a Lorentz-invariant theory.

\begin{figure}
    \centering
    \begin{subfigure}{0.48\textwidth}
        \centering
        \adjustbox{margin=0 0 0 10pt}{\includegraphics[width=7.2cm]{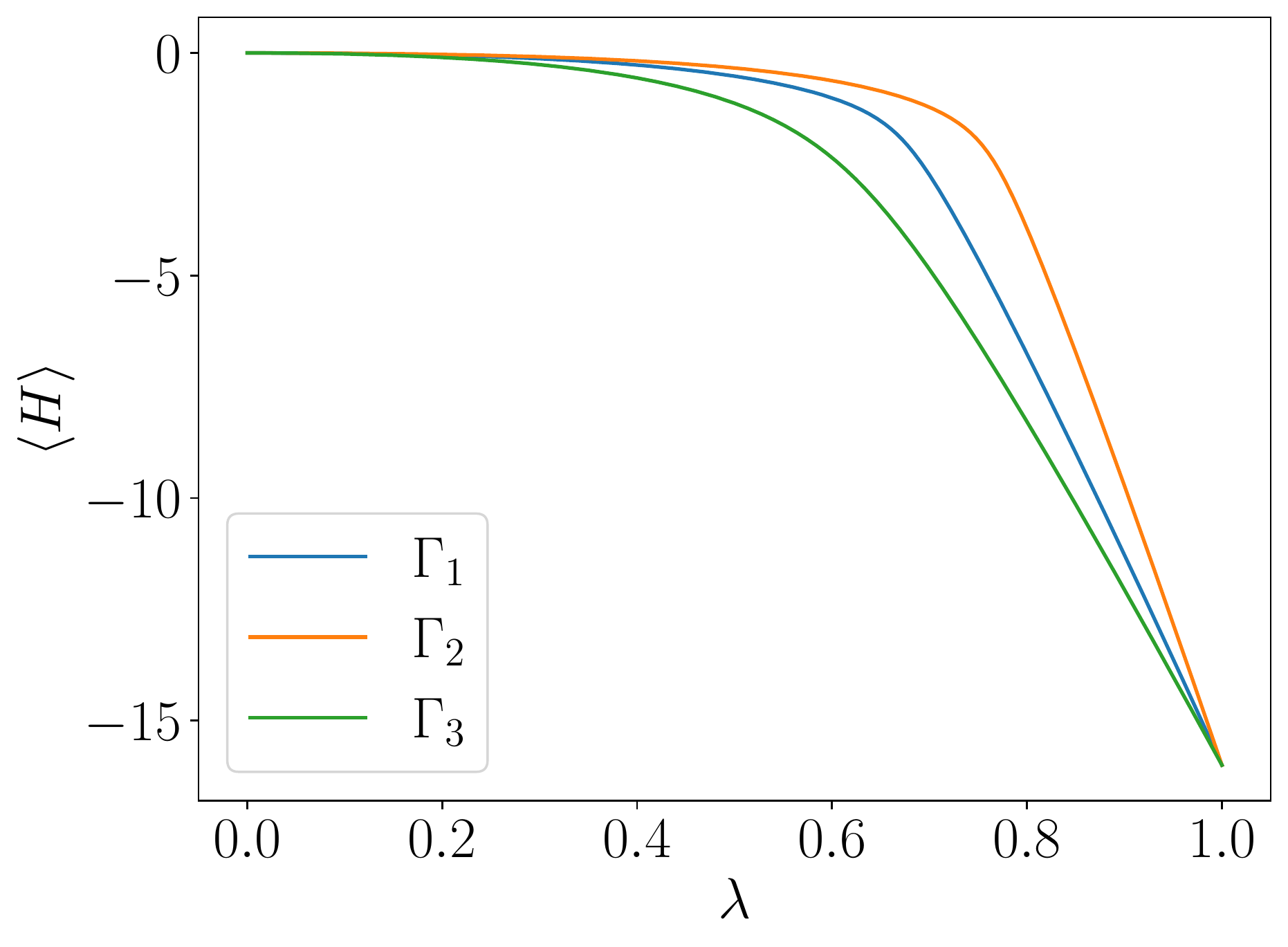}}
        \caption{Ground state energy}
        \label{fig:ground state energy}
    \end{subfigure}
    \quad
    \begin{subfigure}{0.48\textwidth}
        \centering
        \adjustbox{margin=0 0 0 10pt}{\includegraphics[width=7.1cm]{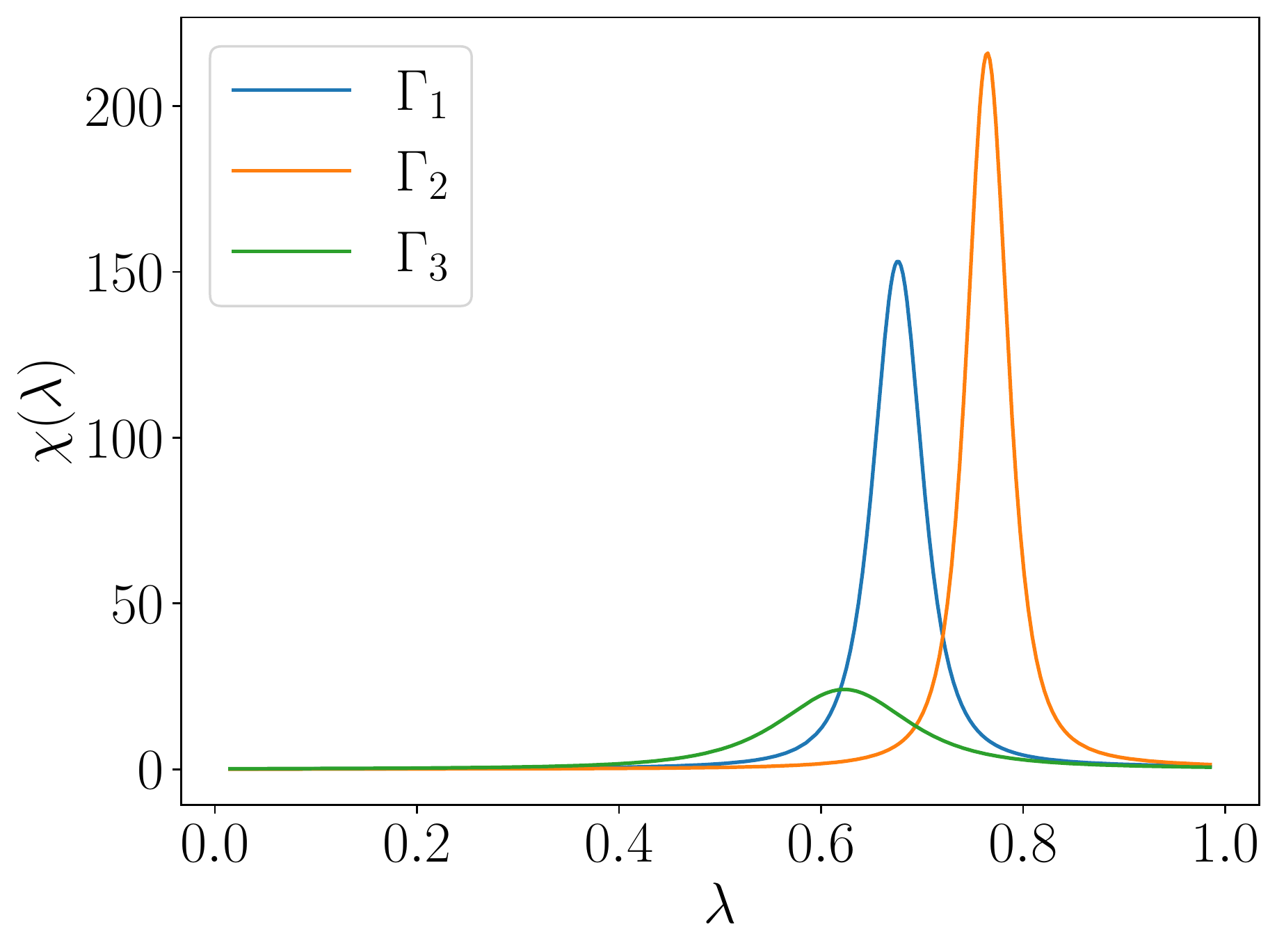}}
        \caption{Fidelity susceptibility}
        \label{fig:fidelity susceptibility}
    \end{subfigure}
    \quad
    \begin{subfigure}{0.48\textwidth}
        \centering
        \adjustbox{margin=0 0 0 10pt}{\includegraphics[width=7.0cm]{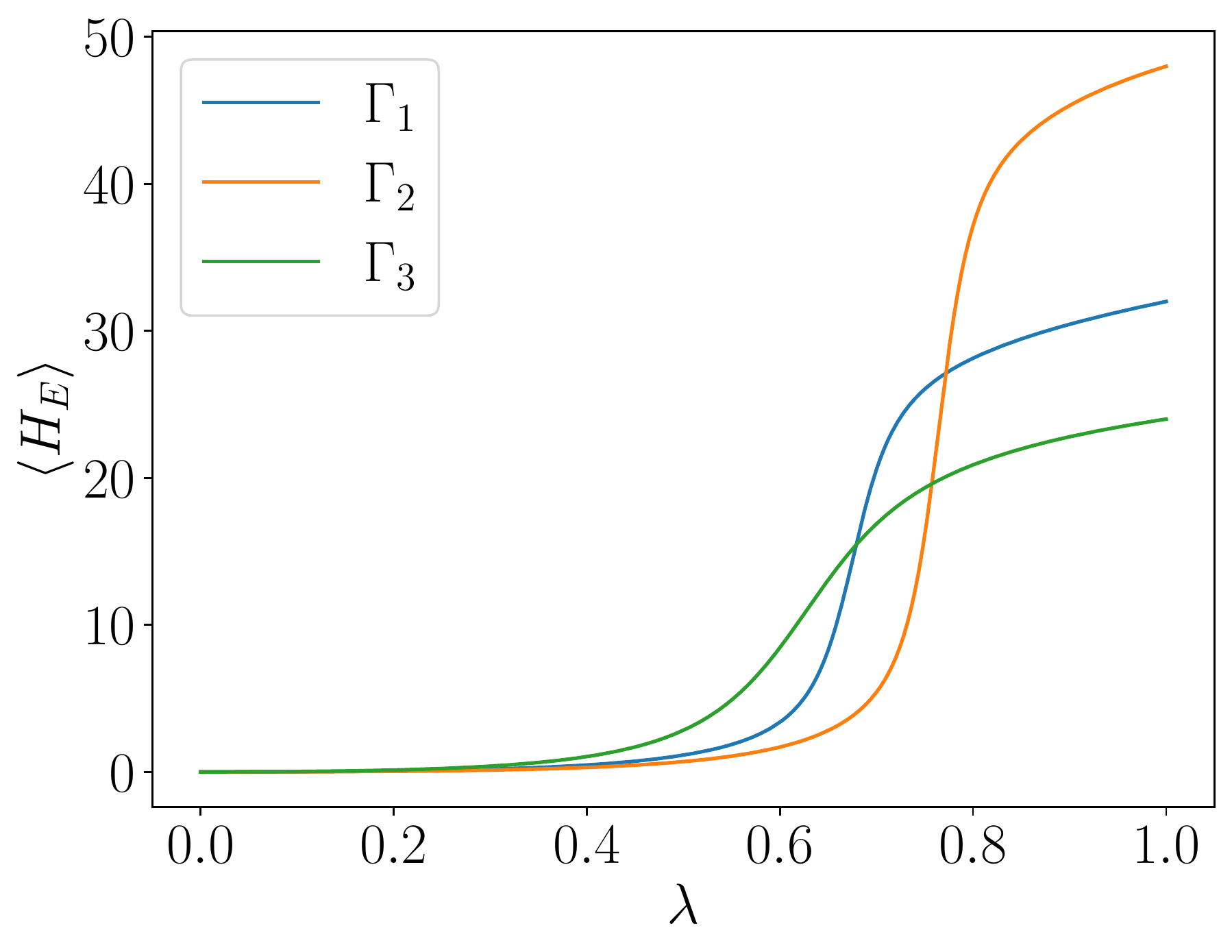}}
        \caption{Ground state $\expval{H_E}$}
        \label{fig:HE expval}
    \end{subfigure}
    \quad
    \begin{subfigure}{0.48\textwidth}
        \centering
        \adjustbox{margin=0 0 0 10pt}{\includegraphics[width=7.1cm]{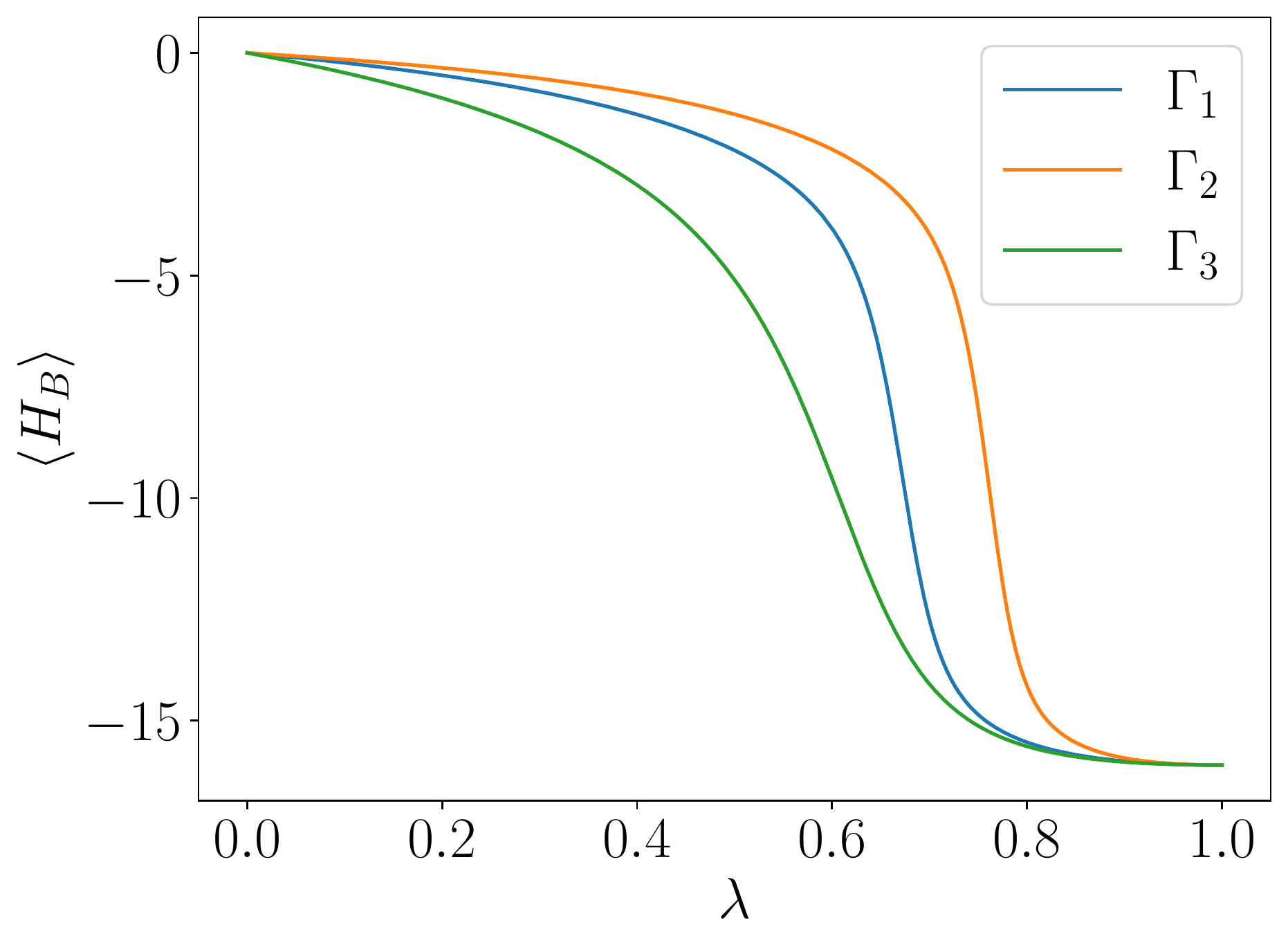}}
        \caption{Ground state $\expval{H_B}$}
        \label{fig:HB expval}
    \end{subfigure}
    \quad
    \begin{subfigure}{0.48\textwidth}
        \centering
        \adjustbox{margin=0 0 0 10pt}{\includegraphics[width=7.2cm]{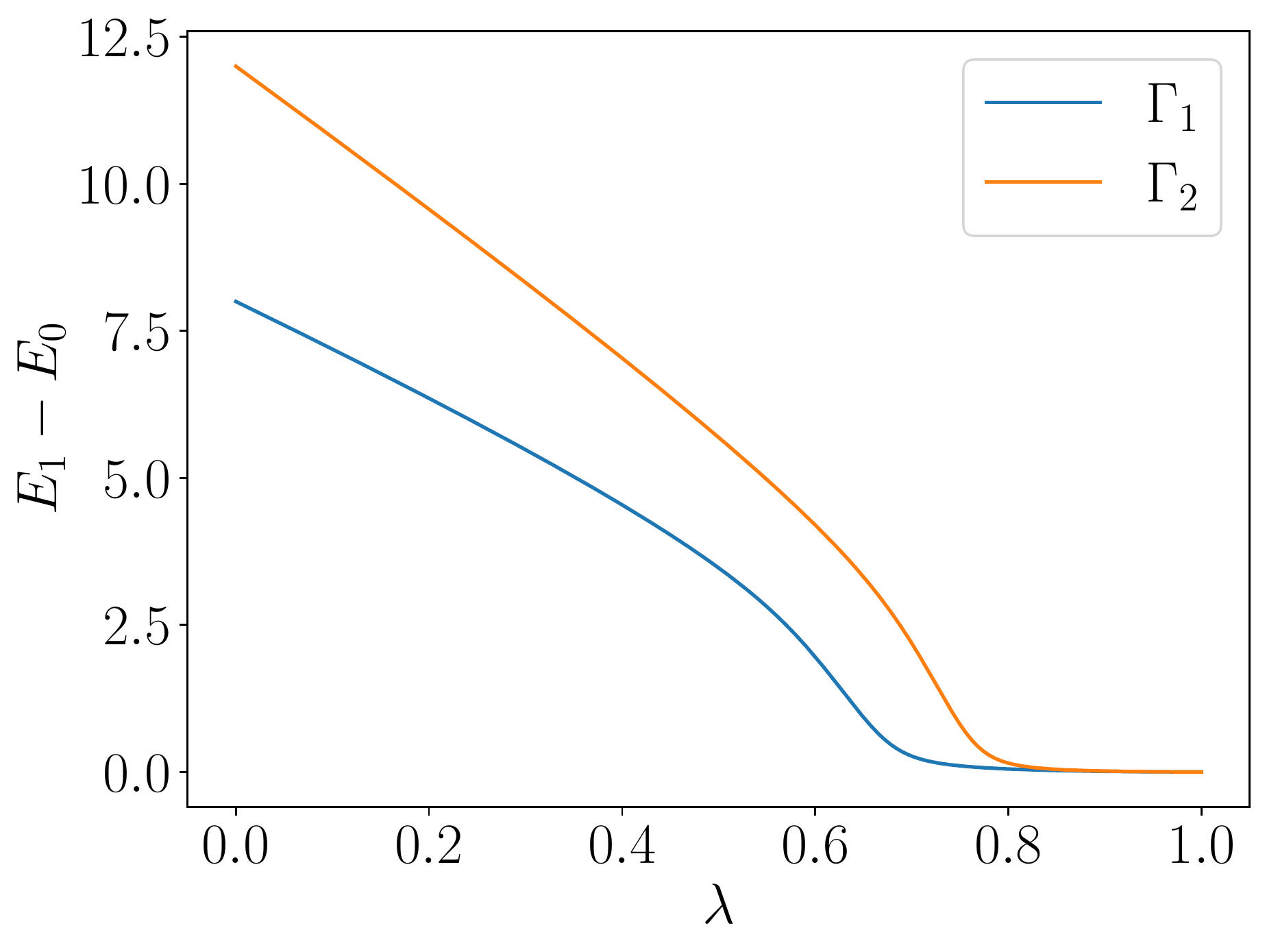}}
        \caption{$E_1-E_0$ for $\Gamma_1$ and $\Gamma_2$}
        \label{fig:energy gap Gamma_1 Gamma_2}
    \end{subfigure}
    \quad
    \begin{subfigure}{0.48\textwidth}
        \centering
        \adjustbox{margin=0 0 0 10pt}{\includegraphics[width=7.1cm]{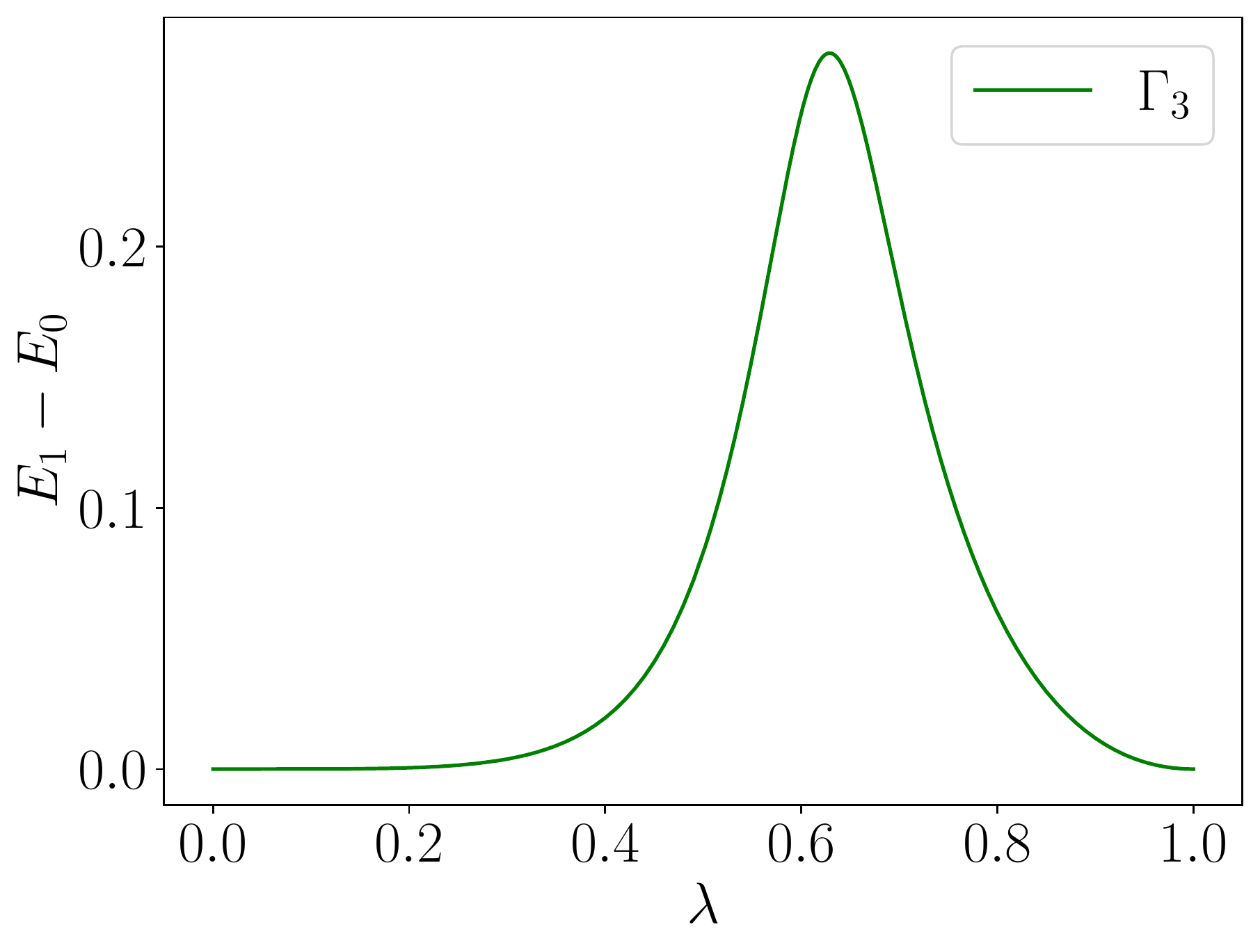}}
        \caption{$E_1-E_0$ for $\Gamma_3$}
        \label{fig:energy gap Gamma_3}
    \end{subfigure}
    \quad
    \caption{Exact diagonalization results for $D_4$ gauge theory on a $2 \times 2$ periodic lattice in the gauge-invariant basis, for three different choices of generating set $\Gamma$, i.e. $\Gamma_1 = \{r,r^3,s,r^2s\}$ (in blue), $\Gamma_2=\{r, r^3, s, rs, r^2s, r^3s\}$ (relativistic, in orange) and $\Gamma_3 = \{r, r^2, r^3\}$ (degenerate, in green). (a): ground state energy. (b): fidelity susceptibility. (c-d): ground state expectation values of the electric and magnetic Hamiltonians. (e-f): energy difference between the first two energy levels. }
    \label{fig:numerical results}
\end{figure}

The results of the exact diagonalization are shown in Fig.~\ref{fig:numerical results}. The ground state energy, and the ground state expectation values of the electric and magnetic Hamiltonians are shown in Figs.~\ref{fig:ground state energy}, \ref{fig:HE expval}, \ref{fig:HB expval} respectively.
The qualitative picture is rather similar in all three cases, with the electric energy increasing with $\lambda$ and the magnetic energy decreasing with $\lambda$.
The ground state energies coincide at $\lambda = 0$, because $h_E$ always has a zero eigenvalue, and at $\lambda=1$ where the Hamiltonian reduces to $H_B$, which is the same in the three cases. 
The plaquette Wilson loop is equal to $H_B$ apart from an overall prefactor, and therefore its behaviour is also given by Fig.~\ref{fig:HB expval}. We note that our data for the ground state energies agrees with that obtained in \cite{marchesethesis} with a different method.

The difference in energy between the first two states $E_1-E_0$ is shown in Figs.~\ref{fig:energy gap Gamma_1 Gamma_2}, \ref{fig:energy gap Gamma_3}. The difference $E_1-E_0$ coincides with the energy gap except where there are degenerate ground states, in which case $E_1-E_0=0$. Ground state degeneracy occurs in all three cases at $\lambda=1$ and, for $\Gamma_3$, also at $\lambda=0$. For $\Gamma_3$, the ground state degeneracy at $\lambda=0$ is lifted at any $\lambda > 0$ but the scale of the gap remains much smaller than in the other two cases. In all three cases we recognize a transition region at $\lambda \sim 0.6 - 0.8$ (depending on $\Gamma$), in which the ground state quickly evolves from minimizing the electric energy to minimizing the magnetic energy. For $\Gamma_1$ and $\Gamma_2$, at the transition the gap rapidly becomes very small, while for $\Gamma_3$ it is always close to zero except in the transition region, where it peaks.

One possible way to locate the transition point is to identify it as the point of sharpest variation of $\expval{H_E}$ and/or $\expval{H_B}$ (i.e. the maximum of the absolute value of their derivative with respect to $\lambda$). With this identification, the transition points given by either $\expval{H_E}$ or $\expval{H_B}$ coincide at $\lambda_1^* = 0.67(1)$ and $\lambda_2^* = 0.76(1)$ for $\Gamma_1$ and $\Gamma_2$, but show a small difference for $\Gamma_3$, at $\lambda_{3,E}^* = 0.63(1)$ and $\lambda_{3,B}^* = 0.61(1)$. Alternatively, calling $\ket{\psi_0(\lambda)}$ the ground state of $H(\lambda)$, one can look at the fidelity susceptibility \cite{FidelitySusc}
\begin{equation}
    \chi(\lambda) = -\frac{\partial^2}{\partial \epsilon^2} \log \abs{\expval{\psi_0(\lambda) | \psi_0(\lambda+\epsilon) }}^2 \bigg\lvert_{\epsilon=0} \ ,
\end{equation}
which is expected to peak at the transition \cite{FidelitySuscMadeSimple}. Fig.~\ref{fig:fidelity susceptibility} shows $\chi(\lambda)$ for the three cases and its peak identifies the transition point as $\lambda_{1, \chi}^*=0.67(1)$, $\lambda_{2, \chi}^*=0.76(1)$ and $\lambda_{3, \chi}^*=0.62(1)$, in agreement with the previous method.

Overall, these results point towards the expected picture of a two-phase structure for all three cases. The data for $\Gamma_1$ and $\Gamma_2$ are consistent with the usual picture of a confining phase at small $\lambda$ and a deconfined phase at large $\lambda$. For $\Gamma_3$, the gap is much smaller, especially at small $\lambda$, which complicates the interpretation of this phase as confining. Of course, due to the small volume these results are only qualitative and preliminary; a study with larger volumes would be required in order to properly establish the phase structure. However, they point to the possibility that theories with electric degeneracy may display different behaviour and phase structure compared to those with no electric degeneracy.

\section{Conclusions}

In this work we considered Hamiltonians for gauge theories with a finite gauge group and we have shown that the electric term may be interpreted as a natural Laplacian operator on the finite group, constructed as the graph Laplacian of its Cayley graph. The choice of generating set of the Cayley graph has a simple relation with the ground state degeneracy of the electric Hamiltonian. We have also given careful consideration to the various choices involved in constructing a finite group gauge theory and their consequences. Independently from the choice of Hamiltonian, we have shown that the physical, gauge-invariant Hilbert space of pure gauge theories may be explicitly described in terms of spin-network states, which are particularly suitable for finite groups. This also allows us to derive a simple formula to compute the dimension of the physical Hilbert space on an arbitrary lattice. Using the spin-network basis, we diagonalized $D_4$ gauge theory on a small periodic lattice with different Hamiltonians. Due to the small system size, these results are only suggestive, but they point to the possibility that theories with a degenerate electric Hamiltonian may have a different phase structure than commonly expected.

The methods employed in this work may be extended in several directions. The graph Laplacian construction may be adapted to those approaches where a Lie group is discretized to a finite subset, not necessarily a subgroup \cite{Hackett_2019}. In that case the finite subset may be seen as a weighted graph, with the edge weights representing the distance between group elements in the parent Lie group.

As we have seen, working directly in the gauge-invariant basis reduces the size of the Hilbert space and and implements the Gauss' law exactly, at the expense of higher complexity of the Hamiltonian. It would be worthwhile to explore whether the gauge-invariant basis can be efficiently implemented, for example in a quantum circuit. It is also possible to extend the spin-network basis to gauge theories coupled to matter fields, and we will treat this case in a future publication.

Finally, it would be very interesting to explore the possibility of a non-standard mechanism to obtain a continuum limit for finite group gauge theories; for example, this is possible in quantum link models via the D-theory formulation \cite{Brower_2004, WIESE2006336}. 

\section*{Acknowledgments}

The authors would like to thank L. Marchese for providing his data for comparison and G. Kanwar, S. Pascazio, P. Facchi for helpful discussions. AM acknowledges funding from the Schweizerischer Nationalfonds (grant agreement number 200020\_200424). EE and SP are supported by Istituto Nazionale di Fisica Nucleare (INFN) through the project \say{QUANTUM} and the QuantERA 2020 Project \say{QuantHEP}. A subset of this work was published as part of the thesis \cite{marianithesis}.

\appendix

\section{Some groups of interest}\label{sec:some groups of interest}

\subsection{The cyclic groups \texorpdfstring{$\Z_N$}{ZN} gauge theory}\label{sec:ZN groups}

The cyclic group $\Z_N$ is an Abelian group of order $N$. It is generated by one element $\xi$, which thus satisfies $\xi^N=1$. Thus $\Z_N = \{1, \xi, \xi^2, \ldots, \xi^{N-1}\}$. Since $\Z_N$ is Abelian, its conjugacy classes are singlets, i.e. it has $N$ conjugacy classes of one element; moreover, it has $N$ inequivalent irreps, all of which are one-dimensional,
\begin{equation}
    \rho_j (\xi^k) = \omega_N^{kj}  \ , \,\,\,\,\,\,\, j = 0,1,\ldots, N-1 \ ,
\end{equation}
with $\omega_N= e^{2\pi i /N}$. The bases $\{\ket{\xi^k}\}$ and $\{\ket{j}\}$ are related by
\begin{equation}
    \ket{j} = \sum_{k=0}^{N-1} \braket{\xi^k|j} \ket{\xi^k} = \frac{1}{\sqrt{N}}\sum_{k=0}^{N-1} \omega_N^{kj} \ket{\xi^k} \ ,
\end{equation}
which is just the discrete Fourier transform. 

\subsection{The dihedral groups \texorpdfstring{$D_N$}{DN}}\label{sec:DN groups}

The dihedral groups $D_N$ are non-Abelian groups of order $2N$. They are subgroups of $\Or(2)$, and they are generated by a rotation $r$ and a reflection $s$, which satisfy $r^N=s^2=1$ and $srs = r^{-1}$. We describe in more detail the dihedral group $D_4$ of order $8$. Its elements are $D_4 = \{1, r, r^2, r^3, s, r s, r^2 s, r^3 s \}$ and it has $5$ conjugacy classes, $\{1\},\{r,r^3\},\{r^2\},\{s,r^2s\},\{rs,r^3s\}$. It has $5$ irreducible representations, which we number from $j=0$ (trivial representation) to $j=4$. All the irreps are one-dimensional except $j=4$, which is two-dimensional and faithful. The character table is shown in Table \ref{tab:char}.

\begin{table}[t]
    \centering
    \begin{tabular}{cccccc}
        \toprule
         &  $\{1\}$ & $\{r,r^3\}$ & $\{r^2\}$ & $\{s,r^2s\}$ & $\{rs,r^3s\}$ \\
        \midrule
        $\chi_0$ & $+1$ & $+1$ & $+1$  & $+1$ & $+1$ \\
        $\chi_1$ & $+1$ & $-1$ & $+1$  & $+1$ & $-1$ \\
        $\chi_2$ & $+1$ & $+1$ & $+1$  & $-1$ & $-1$ \\
        $\chi_3$ & $+1$ & $-1$ & $+1$  & $-1$ & $+1$ \\
        $\chi_4$ & $+2$ & $0$  & $-2$  & $0$  & $0$ \\
        \bottomrule
    \end{tabular}
    \caption{Character table of $D_4$}
    \label{tab:char}
    \vspace{-3mm}
\end{table}

As the $j=4$ is the only faithful representation, it is a natural choice for the magnetic Hamiltonian.

\section{Proofs}

\subsection{Degeneracy of electric Hamiltonian}\label{sec:laplacian degeneracy}

As discussed in Section \ref{sec:hamiltonian formulation}, the degeneracy of the electric Hamiltonian given by the finite group Laplacian $\Delta$ is directly related to the structure of the Cayley graph. In particular, it is a standard result that the graph Laplacian always has a zero mode and its degeneracy equals the number of connected components of the graph \cite{spectralgraphtheory}. Here we show that the Cayley graph is connected if and only if its generating set $\Gamma$ generates the whole group. If instead $\expval{\Gamma} \neq G$, then the Cayley graph splits into connected components identified with the cosets of $\expval{\Gamma}$ in $G$; thus the degeneracy of the finite-group Laplacian $\Delta$ equals $\abs{G}/\abs{\expval{\Gamma}}$.

Any subset $\Gamma \in G$ generates a subgroup $\expval{\Gamma} < G$. The right cosets of $\expval{\Gamma}$ are of the form $\expval{\Gamma} h$ for $h$ in $G$. Since cosets partition the group, any two group elements $g_1$ and $g_2$ will belong to some coset, say $g_1 \in \expval{\Gamma} h_1$ and $g_2 \in \expval{\Gamma} h_2$. We want to show that there is an edge in the Cayley graph between group elements $g_1$ and $g_2$ if and only if they are in the same coset, i.e. $\expval{\Gamma} h_1 = \expval{\Gamma} h_2$. The fact that $g_i \in \expval{\Gamma} h_i$ means that $g_i = k_i h_i$ for some $k_i \in \expval{\Gamma}$. Moreover, there is an edge between $g_1$ and $g_2$ if and only if $g_1 g_2^{-1} = k_1 h_1 h_2^{-1} k_2 \in \Gamma$. But since $k_i \in \expval{\Gamma}$ this is equivalent to saying that $h_1 h_2^{-1} \in \expval{\Gamma}$, which is equivalent to $\expval{\Gamma} h_1 = \expval{\Gamma} h_2$. This concludes the proof. 

\subsection{Counting of invariant states}\label{sec:counting invariant states}

In Section \ref{eq:physical hilbert space dimension} we used the fact that for a generic representation $\rho$, the dimension of the space of invariant vectors is given by
\begin{equation}
    \dim{\mathrm{Inv}(\rho)} = \frac{1}{\abs{G}} \sum_{g \in G} \chi_\rho(g) \ .
\end{equation}
If $\rho$ is a non-trivial irreducible then the corresponding character sums to zero \cite{Serre} and there are no invariant states. This is to be expected since, by definition, irreducible representations have no non-trivial invariant subspaces, but any invariant vector would span an invariant subspace.

Here we provide a proof of the above formula. If $v$ is an invariant vector for the representation $\rho$, by definition it satisfies $\rho(g) v = v$ for all $g \in G$. Now we construct a projector onto the subspace of invariant vectors. We define the averaging map $\mathrm{Av}: V_\rho \to V_\rho$,
\begin{equation}
    \mathrm{Av}(v) = \frac{1}{\abs{G}} \sum_{g \in G} \rho(g) v \ .
\end{equation}
The averaging map is the projector onto the subspace of invariant vector. In fact, given an arbitrary vector $v$, we see that $\mathrm{Av}(v)$ is invariant because
\begin{equation}
    \rho(g) \mathrm{Av}(v) = \frac{1}{\abs{G}} \sum_{h \in G} \rho(g h) v = \frac{1}{\abs{G}} \sum_{h \in G} \rho(h) v = \mathrm{Av}(v) \ .
\end{equation}
Therefore, $\mathrm{Av}$ maps the representation space to the subspace of invariant vectors $\mathrm{Av}: V_\rho \to \mathrm{Inv}(V_\rho)$. Moreover, if $v$ is invariant, then $\mathrm{Av}(v) = v$, and more generally, $\mathrm{Av}^2 = \mathrm{Av}$ by a similar calculation. This means that $\mathrm{Av}$ is a projector onto the subspace of invariant vectors. As usual, the size of projected subspace is given by the trace of the projector, $\dim{\mathrm{Inv}(\rho)} = \tr{\mathrm{Av}}$, which reproduces the above formula.

\printbibliography

\end{document}